\documentclass[11pt,a4paper]{article}

\usepackage[T1]{fontenc} 
\usepackage{amsmath}
\usepackage{amsthm}
\usepackage{amsfonts}
\usepackage{amssymb}
\usepackage{graphicx,psfrag}
\usepackage{hyperref}
\usepackage[font=footnotesize,labelfont=bf]{caption}
\usepackage{amscd}

\usepackage{enumerate}

\begin{document}
\def\thefootnote{\fnsymbol{footnote}}

\begin{center}
\Large{\textbf{The Dynamical Evolution of 3-Space\\in a Higher Dimensional Steady State Universe}} \\[0.5cm]
 
\large{\"{O}zg\"{u}r Akarsu$^{\rm a,\,b}$, Tekin Dereli$^{\rm a}$}
\\[0.5cm]

\small{
\textit{$^{\rm a}$ Department of Physics, Ko\c{c} University, 34450 Sar{\i}yer, {\.I}stanbul, Turkey}}

\vspace{.2cm}

\small{
\textit{$^{\rm b}$ Abdus Salam International Centre for Theoretical Physics, Strada Costiera 11, 34151, Trieste, Italy}}

\end{center}

\vspace{.6cm}

\hrule \vspace{0.3cm}
\noindent \small{\textbf{Abstract}\\ 
We investigate a class of cosmological solutions of Einstein's field equations in higher dimensions with
a cosmological constant and an ideal fluid matter distribution as a source. We discuss the dynamical evolution of the universe
subject to two constraints that (i) the total volume scale factor of the universe is constant and (ii) the effective energy density is constant. We obtain various interesting new dynamics for the external space that yield a time varying deceleration parameter including oscillating cases when the flat/curved external and curved/flat internal spaces are considered. We also comment on how the universe would be conceived by an observer in four dimensions who is unaware of the extra dimensions.}
\\
\noindent
\hrule
\def\thefootnote{\arabic{footnote}}
\setcounter{footnote}{0}

\let\thefootnote\relax\footnote{\textbf{E-Mail:} oakarsu@ku.edu.tr, tdereli@ku.edu.tr}

\def\thefootnote{\arabic{footnote}}
\setcounter{footnote}{0}

\section{Introduction}
\label{Intro}

The idea that the spacetime has actually more than four dimensions but appears to be four dimensional because the extra space dimensions are too small for local detection goes directly back to the years after Einstein's general relativitistic theory of gravitation was first introduced in 1915. Kaluza and Klein's attempt to unify gravitation and electromagnetism during the 1920s was based on the assumption that the universe we live in is in fact five dimensional, but as the fifth dimension remains small, it appears effectively four dimensional \cite{OverduinWesson97}. The unification of fundamental interactions of nature achieved in higher dimensions in more recent years provides a strong motivation to give a serious consideration to this possibility. We know today that anomaly-free superstring models of all fundamental interactions require a spacetime of ten dimensions for consistency and the M-theory in which they are embedded lives in an eleven dimensional spacetime (see \cite{Lidsey00} and references therein). It is generally assumed that all but four of the spacetime dimensions are compactified on an unobservable internal manifold, leaving an observable $(1+3)$-dimensional spacetime. On the other hand, in the early 1980s the dynamical (cosmological) reduction of internal dimensions to unobservable scales with the external physical dimensions expanding while the internal dimensions contracting started to be considered in a cosmological context \cite{ChodosDetweiler80,Freund82,DereliTucker83}. Since then different possibilities have been investigated: Cosmological models where the external dimensions are expanding while the internal dimensions are i) contracting \cite{ChodosDetweiler80,Freund82,DereliTucker83}, ii) static \cite{Barrow86,Bringmann03}, iii) expanding at a much slower rate than the external dimensions \cite{AkarsuDereli12} or iv) exhibiting any combination of these possibilities \cite{ChodosDetweiler80,BleyerZhuk96}.

The accelerating expansion of the universe which is one of the crucial components of contemporary cosmology has further reinforced the interest in higher dimensional cosmological models. An accelerating expansion phase of the universe first came into serious consideration when it was realized that the presence of an accelerating expansion epoch in the early universe (inflation) could resolve the problems of standard Big Bang cosmology such as the observed spatial homogeneity, isotropy and flatness of the universe (e.g., \cite{Guth81,Linde82}). 
Today it is widely believed that the inflationary scenario is one of the most prominent attempts to resolve the problems of standard Big Bang cosmology. Surprisingly, the direct observational realization of the possibility of an accelerating expansion of the universe came from the discovery of the current acceleration of the universe from SNIa observations \cite{Riess98,Perlmutter99} and is supported by independent observations e.g. the WMAP observations \cite{Komatsu11}. However, most of the inflationary scenario variations are based on general relativity and the accelerating expansion is generated by \textit{ad hoc} scalar fields that can behave like a positive cosmological constant under some conditions and there is not yet a concrete derivation of inflation from a fundamental theory such as the string theory \cite{Quevedo}. On the other hand, the current acceleration of the universe can be easily explained by the inclusion of a positive cosmological constant into the Einstein's field equations in the presence of pressure-less matter. This is the starting point of $\Lambda$CDM cosmology, which is the simplest model that can accommodate the observed dynamics of the universe. However, a positive cosmological constant is mathematically equivalent to a vacuum energy density and the theoretical estimates for it exceed the observational limits by some 120 orders of magnitude. This is still one of the most pressing unsolved problems in fundamental physics \cite{Zeldovich,Weinberg89,Sahni00} and it led to the dark energy concept \cite{Copeland06} in cosmology. To address these issues the presence of higher dimensions has also been considered among several other possibilities such as modifying general relativity or predicting the presence of unknown energy sources, i.e. inflaton and dark energy fields. The idea is that both the extra dimensions (although they are today out of our reach) and the possible higher dimensional fluids can affect the dynamics of the observed universe and hence can appear as unknown sources in the four dimensional universe where it is interpreted in terms of the conventional general relativity.

There are neither a priori nor observational reasons for assuming that the universe during its dynamical evolution has always been four dimensional and the presence of extra dimensions seems a good place for seeking answers to many questions in cosmology. Indeed, there is a wide literature on higher dimensional cosmological models that have been studied extensively within various approaches and contexts (see for instance references \cite{Sahdev84,Sato84,Ishihara84,Demaret85,Okada86,Leon93,Yearsley96,Mohammedi02,Carroll09,Reyes11}). In the literature there is an interesting class of models where the total volume of the higher dimensional space remains constant \cite{Freund82,DereliTucker83,BleyerZhuk96,RainerZhuk00,HoKephart10}. In particular, Dereli and Tucker \cite{DereliTucker83}, considering the inflationary model introduced by Guth \cite{Guth81}, obtained a higher dimensional general relativistic cosmological model within the framework of flat external and internal spaces by assuming that the external space expands exponentially and the internal space contracts exponentially. It is found that the volume of the higher dimensional space is constant and additionally that the energy density of the higher dimensional fluid is also constant.

A higher dimensional space with a constant volume lives forever with no beginning and no end. Yet, the three dimensional space that is inferred from this eternal higher dimensional space may have dynamics consistent with the observed universe. For instance, it may exhibit the dynamics similar to the standard Big Bang or $\Lambda$CDM model. The constancy of the higher dimensional volume also assures the dynamical contraction (reduction) of the internal space for an expanding three dimensional external space. The three dimensional space can still commence with a coordinate singularity and the size of the higher dimensional space can be chosen so that the physical processes do occur the same as in the four dimensional universe. Another interesting feature of the idea of a constant higher dimensional volume comes out when it is further assumed that the energy density of the higher dimensional effective fluid is also constant. In such a cosmological model, there would occur a mass leakage from the contracting space into the expanding space, say, from the contracting $n$-dimensional space into the expanding $3$-dimensional space. Hence matter is neither created nor exhausted in such a model but is redistributed between the external and internal spaces. The effective $(1+3)$-dimensional spacetime may simulate a steady-state cosmology with a natural mechanism for maintaining a constant density of matter without modifying the mathematical structure of general relativity as assumed in the conventional steady-state cosmological model with continuos creation of matter \cite{BondiGold,HoyleNarlikar}.

In what follows, we study the dynamics of a higher dimensional cosmological model that yields a constant higher dimensional volume and energy density in the context of Einstein's general relativity with a cosmological constant. We obtain various solutions and also discuss the apparent universe for an observer living in four dimensions unaware of the presence of higher dimensions considering these solutions.

\section{The Model}
\label{sec2}

The theory of gravitation we consider is the generalization of the conventional $(1+3)$-dimensional Einstein's gravity to $(1+3+n)$-dimensions in the presence of a $(1+3+n)$-dimensional cosmological constant with a negative sign that is consistent with anti-de Sitter (AdS) vacua:
\begin{equation}
\label{eqn:EFE}
R_{\mu\nu}-\frac{1}{2}Rg_{\mu\nu}+\Lambda g_{\mu\nu} =-kT_{\mu\nu}.
\end{equation}
Here the indices $\mu$ and $\nu$ run through $0,1,2,...,3+n$ and $0$ is reserved for the time coordinate $t$. $R_{\mu\nu}$, $R$ and $g_{\mu\nu}$ are the Ricci tensor, Ricci scalar and the metric tensor, respectively, of the $(1+3+n)$-dimensional spacetime. $T_{\mu\nu}$ is the energy-momentum tensor of a $(1+3+n)$-dimensional inviscid, incompressible fluid and $k=8\pi G$ where $G$ is the $(1+3+n)$-dimensional gravitational coupling constant.

We consider $(1+3+n)$-dimensional spacetime manifold with product topology
\begin{equation}
\mathcal{M}^{1+3+n}=\mathcal{R}\times \mathcal{M}^{3}\times \mathcal{K}^{n},
\end{equation}
where $\mathcal{R}$ is the manifold of time, $\mathcal{M}^3$ is the manifold of $3$-dimensional external space that represents the space we observe and $\mathcal{K}^{n}$ is the manifold of the $n$-dimensional compact internal space that may be so small to be observed directly and locally. We define, on this manifold, a spatially homogenous but not necessarily isotropic $(1+3+n)$-dimensional synchronous spacetime metric that involves $3$-dimensional external space with constant curvature for $\mathcal{M}^3$ and $n$-dimensional internal space with constant curvature for $\mathcal{K}^n$:
\begin{eqnarray}
\label{eqn:metric}
ds^2=-dt^2+a^2(t)\,\frac{dx_{1}^{2}+dx_{2}^{2}+dx_{3}^{2}}{\left[1+\frac{\kappa_{a}}{4}(x_{1}^{2}+x_{2}^{2}+x_{3}^{2})\right]^2}
+ s^2(t)\,\frac{ dy_{1}^{2} +...+ dy_{n}^{2}}{\left[1+\frac{\kappa_{s}}{4}(y_{1}^{2}+...+y_{n}^{2})\right]^2}.
\end{eqnarray}
Here $a(t)$ is the scale factor, $\kappa_{a}$ is the curvature index of the 3-dimensional external space. $s(t)$ is the scale factor, $\kappa_{s}$ is the curvature index and $n=1,2,3,...$ is the number of the internal dimensions of the $n$-dimensional internal space. The curvature indices $\kappa_{a}$ and $\kappa_{s}$ can take values $-1$, $0$ and $1$. We note that the manifold $\mathcal{R}\times \mathcal{M}^{3}$ is equipped with the usual Robertson-Walker metric and $\mathcal{M}^{3}$ can be either open, flat or closed that correspond to $\kappa_{a}=-1,\,0$ and $1$, respectively. On the other hand, it is important that the internal space is compact since we require internal space to yield finite volume. It is known that there are non-trivial global topologies that are compact for any sign of the sectional curvature. In the case $\kappa_{s}=1$ the $n$-dimensional space is $n$-sphere ($\mathcal{K}^{n}=\mathcal{S}^{n}$), in the case $\kappa_{s}=0$ the most simple example for a compact $n$-dimensional space is the $n$-dimensional torus ($\mathcal{K}^{n}=\mathcal{T}^{n}$). We note that we also consider negative constant sectional curvature $\kappa_{s}=-1$ (hyperbolic geometry) for the internal space. Such spaces are also compact if they have a quotient structure such that $\mathcal{K}^{n}=\mathcal{H}^{n}/\Gamma$, where $\mathcal{H}^{n}$ and $\Gamma$ are $n$-dimensional hyperbolic space and its discrete isometry group, respectively \cite{Ratcliffe06}. In the following we shall refer to the cases $\kappa_{s}=-1,\,0$ and $1$ as the open, flat and closed internal space in accordance with the common usage in cosmology.

We describe the $(1+3+n)$-dimensional fluid with an energy-momentum tensor that yields distinct pressures in the external and internal spaces:
\begin{equation}
\label{eqn:EMT}
{T_{\mu}}^{\nu}={\textnormal{diag}}[-\rho, p_{{\rm{e}}},p_{{\rm{e}}},p_{{\rm{e}}},p_{{\rm{i}}},...,p_{{\rm{i}}}]
\end{equation}
where $\rho$ is the energy density, $p_{{\rm{e}}}$ and $p_{{\rm{i}}}$ are the pressures that are associated with the external and internal dimensions, respectively. In fact, this is the most general form of the energy-momentum tensor that can be used for describing a fluid at rest in comoving coordinates, i.e. whose $(1+3+n)$-velocity is $u^{\mu}=(1,0,0,...)$, within the framework of the spacetime defined by the metric (\ref{eqn:metric}). Since we do not know the nature of the physical ingredients of the higher dimensional universe, we conveniently allow the possibility of an energy-momentum tensor with distinct and dynamical pressures in the external and internal spaces.

Einstein's field equations (\ref{eqn:EFE}) for the metric (\ref{eqn:metric})  in the presence of the energy-momentum tensor given by (\ref{eqn:EMT}) lead to the following system of differential equations:
\begin{equation}
\label{eqn:EFE1}
3\frac{\dot{a}^2}{a^2}+3\frac{\kappa_{a}}{a^2}+3n\frac{\dot{a}}{a}\frac{\dot{s}}{s}+\frac{1}{2}n(n-1)\frac{\kappa_{s}+\dot{s}^2}{s^2}+\Lambda=k\rho,
\end{equation}
\begin{eqnarray}
\label{eqn:EFE2}
\frac{\dot{a}^2}{a^2}+2\frac{\ddot{a}}{a}+\frac{\kappa_{a}}{a^2}+n\frac{\ddot{s}}{s}+2n\frac{\dot{a}}{a}\frac{\dot{s}}{s}+\frac{1}{2}n(n-1)\frac{\kappa_{s}+\dot{s}^2}{s^2}+\Lambda = -kp_{{\rm{e}}},
\end{eqnarray}
\begin{eqnarray}
\label{eqn:EFE3}
3\frac{\dot{a}^2}{a^2}+3\frac{\ddot{a}}{a}+3\frac{\kappa_{a}}{a^2}+(n-1)\frac{\ddot{s}}{s}+3(n-1)\frac{\dot{a}}{a}\frac{\dot{s}}{s}+\frac{1}{2}(n-1)(n-2)\frac{\kappa_{s}+\dot{s}^2}{s^2}+\Lambda = -kp_{{\rm{i}}}.
\end{eqnarray}
From the Bianchi identity, we know that Einstein's field equations imply the conservation of the energy-momentum tensor, i.e. the continuity equation for the higher dimensional fluid follows:
\begin{equation}
\label{eqn:bianchi}
\dot{\rho}+\left(3\frac{\dot{a}}{a}+n\frac{\dot{s}}{s}\right)\rho+3\frac{\dot{a}}{a}p_{e}+n\frac{\dot{s}}{s}p_{i}=0.
\end{equation}
The field equations (\ref{eqn:EFE1})-(\ref{eqn:EFE3}) should be satisfied by five unknown functions $a$, $s$, $\rho$, $p_{{\rm{e}}}$ and $p_{{\rm{i}}}$ and therefore the system is not fully determined. We must provide further constraint equations for a full determination. Here, we are interested in higher dimensional cosmologies that exhibit the following two properties:
\begin{enumerate}[(i)]
\item
The higher dimensional universe has a constant volume as a whole but the internal and external spaces are dynamical.
\item
The energy density is constant in the higher dimensional universe.
\end{enumerate}
In accordance with these requirements, we assume that the $(3+n)$-dimensional \textit{volume scale factor} of the universe is constant:
\begin{equation}
\label{eqn:ansatz1}
V_{3+n}\equiv a^{3}s^{n}=V_{0}=\textnormal{constant}.
\end{equation}
Note that although the higher dimensional volume scale factor $V_{3+n}$ is constant and finite, the volume scale factors of the external space $V_{3}=a^{3}$ and that of the internal space $V_{n}=s^{n}$ are not necessarily constant and finite. The above condition also assures the dynamical contraction (hence reduction) of the internal space for an expanding external space. As a second constraint we assume that the energy (mass) density of the $(1+3+n)$-dimensional matter is a positive definite constant:
\begin{equation}
\label{eqn:ansatz2}
\rho=\rho_{0}>0.
\end{equation}
This means that matter need not be created nor exhausted in the higher dimensional universe but will be redistributed between the external and internal spaces as the time progresses. There should be a matter leakage from the contracting space into the expanding space, that mimics a continuos matter creation in the expanding space that leads to a modification in general relativity, just as it is in the four dimensional steady-state model introduced by Bondi and Gold \cite{BondiGold} in 1948.
\\
\linebreak 
With the above assumptions, our system becomes fully determined and we can proceed with an exact solution. Using the constraints (\ref{eqn:ansatz1}) and (\ref{eqn:ansatz2}), the system of equations (\ref{eqn:EFE1})-(\ref{eqn:EFE3}) are further reduced and can be written in terms of the scale factor of the external space $a$ only:
\begin{equation}
\label{eqn:EFE1r}
-\frac{3}{2}\left(\frac{3+n}{n}\right)\frac{\dot{a}^2}{a^2}+\frac{1}{2}\kappa_{s} n(n-1) V_{0}^{-\frac{2}{n}}a^{\frac{6}{n}}+3\frac{\kappa_{a}}{a^2}+\Lambda=k\rho_{0},
\end{equation}
\begin{equation}
\label{eqn:EFE2r}
\frac{1}{2}\left(\frac{9+5n}{n}\right)\frac{\dot{a}^2}{a^2}-\frac{\ddot{a}}{a}+\frac{1}{2}\kappa_{s} n(n-1) V_{0}^{-\frac{2}{n}}a^{\frac{6}{n}}+\frac{\kappa_{a}}{a^2}+\Lambda=-kp_{{\rm{e}}},
\end{equation}
\begin{equation}
\label{eqn:EFE3r}
\frac{3}{2}\left(\frac{1+n}{n}\right)\frac{\dot{a}^2}{a^2}+\frac{3}{n}\frac{\ddot{a}}{a}+\frac{1}{2}\kappa_{s} (n-1)(n-2) V_{0}^{-\frac{2}{n}}a^{\frac{6}{n}}+3\frac{\kappa_{a}}{a^2}+\Lambda=-kp_{{\rm{i}}}.
\end{equation}
Using the constraints (\ref{eqn:ansatz1}) and (\ref{eqn:ansatz2}) in the continuity equation (\ref{eqn:bianchi}), we find that the pressures in the external and internal spaces should be identical:
\begin{equation}
p_{\rm e}=p_{\rm i}\equiv p.
\end{equation}
Hence, equations (\ref{eqn:EFE2r}) and (\ref{eqn:EFE3r}) are the same and all we need to do is to solve the following system of equations:
\begin{equation}
\label{eqn:EFE1rr}
-\frac{3}{2}\left(\frac{3+n}{n}\right)\frac{\dot{a}^2}{a^2}+\frac{1}{2}\kappa_{s} n(n-1) V_{0}^{-\frac{2}{n}}a^{\frac{6}{n}}+3\frac{\kappa_{a}}{a^2}+\Lambda=k\rho_{0},
\end{equation}
\begin{equation}
\label{eqn:EFE2rr}
\frac{1}{2}\left(\frac{9+5n}{n}\right)\frac{\dot{a}^2}{a^2}-\frac{\ddot{a}}{a}+\frac{1}{2}\kappa_{s} n(n-1) V_{0}^{-\frac{2}{n}}a^{\frac{6}{n}}+\frac{\kappa_{a}}{a^2}+\Lambda = -k p.
\end{equation}
Since $\rho_{0}$ is a positive constant, we first solve (\ref{eqn:EFE1rr}) for the scale factor of the external space $a$ explicitly and then obtain the scale factor of the internal space $s$ by substituting $a$ into (\ref{eqn:ansatz1}). Finally, $p$ can be obtained by substituting $a$ into (\ref{eqn:EFE2rr}). We were not able to obtain the most general analytic solution, i.e., for the case $\kappa_{a}\neq 0$ and $\kappa_{s}\neq 0$. We discuss the following three special cases:
\begin{enumerate}[(i)]
\item 
Both of the external and internal spaces are flat; $\kappa_{a}=\kappa_{s}=0$. This is the solution found in Ref. \cite{DereliTucker83}. This case is given in section \ref{subsec:flatflat}.
\item
The external space is curved but internal space is flat; $\kappa_{a}\neq 0$ and $\kappa_{s}= 0$. One may observe that, in the field equations (\ref{eqn:EFE1rr})-(\ref{eqn:EFE2rr}), for sufficiently small $a$ (i.e. large $s$) values, the term containing the curvature of the internal space would be negligible, while the term containing the curvature of the external space would be important. Hence, this class approximates the general solution for sufficiently small $a$ (i.e. large $s$) values. This case is given in section \ref{subsec:curvedflat}.
\item
The external space is flat but internal space is curved; $\kappa_{a}= 0$ and $\kappa_{s}\neq 0$. Similar to the case (ii), for sufficiently large $a$ (i.e. small $s$) values, the term containing the curvature of the external space now is negligible, while the term containing the curvature of the internal space is important. Hence, this class approximates the general solution for sufficiently large $a$ (i.e. small $s$) values. This case is given in section \ref{subsec:flatcurved}.
\end{enumerate}

\section{Solutions}
\label{solutions}

\subsection{Solution for flat external and flat internal spaces ($\kappa_{a}=\kappa_{s}=0$)}
\label{subsec:flatflat}

Setting $\kappa_{a}=\kappa_{s}=0$, the field equations (\ref{eqn:EFE1rr}) and (\ref{eqn:EFE2rr}) read
\begin{equation}
\label{eqn:EFE1rI}
-\frac{3}{2}\left(\frac{3+n}{n}\right)\frac{\dot{a}^2}{a^2}+\Lambda=k\rho_{0},
\end{equation}
\begin{equation}
\label{eqn:EFE2rI}
\frac{1}{2}\left(\frac{9+5n}{n}\right)\frac{\dot{a}^2}{a^2}-\frac{\ddot{a}}{a}+\Lambda=-kp .
\end{equation}
We determine from (\ref{eqn:EFE1rI}) the cosmological parameters of the external space; the scale factor, Hubble parameter and deceleration parameter, respectively, as follows:
\begin{eqnarray}
\label{eqn:ffa}
a=a_{0}e^{\frac{\bar{n}}{3}\sqrt{\Lambda-k\rho_{0}} \; t},
\end{eqnarray}
\begin{eqnarray}
H_{a}\equiv \frac{\dot{a}}{a}=\frac{\bar{n}}{3}\sqrt{\Lambda-k\rho_{0}},
\end{eqnarray}
\begin{eqnarray}
q_{a}\equiv -\frac{\ddot{a}a}{{\dot{a}}^2}=-1.
\end{eqnarray}
Here we introduce a new parameter $\bar{n}=\sqrt{\frac{6n}{3+n}}$ that takes values in a narrow range according to the number of internal dimensions: for $n=1,2,..,\infty$ we have $\bar{n}=\sqrt{\frac{3}{2}},\sqrt{\frac{12}{5}},...,\sqrt{6}$. Then substituting (\ref{eqn:ffa}) into (\ref{eqn:ansatz1}), we obtain the cosmological parameters of the internal space; the scale factor, Hubble parameter and deceleration parameter, respectively, as follows:
\begin{eqnarray}
s=s_{0}e^{-\frac{\bar{n}}{n}\sqrt{\Lambda-k\rho_{0}} \; t},
\end{eqnarray}
\begin{eqnarray}
H_{s}\equiv \frac{\dot{s}}{s}=-\frac{\bar{n}}{n}\sqrt{\Lambda-k\rho_{0}},
\end{eqnarray}
\begin{eqnarray}
q_{s}\equiv -\frac{\ddot{s}s}{{\dot{s}}^2}=-1 .
\end{eqnarray}
Here and in the rest of the paper we will use $s_{0}=V_{0}^{\frac{1}{n}}a_{0}^{-\frac{3}{n}}$. Finally, substituting (\ref{eqn:ffa}) into (\ref{eqn:EFE2rI}), the pressure of the higher dimensional effective fluid is found to be
\begin{equation}
p=\rho_{0}-2\frac{\Lambda}{k}.
\end{equation}
The equation of state (EoS) parameter of the fluid then follows:
\begin{equation}
\frac{p}{\rho_{0}}= w=1-2\frac{\Lambda}{k\rho_{0}}.
\end{equation}

It should be observed that, in this model, a consistent dynamical cosmology requires $\Lambda>k\rho_{0}$, and hence $w<-1$. In this model, the external space expands exponentially while the internal space contracts exponentially. Therefore, the external space, i.e., the space we observe, exhibits \textit{de Sitter} expansion. This is the well known solution of the $(1+3)$-dimensional Einstein field equations in the presence of a positive cosmological constant ($\Lambda_{\textnormal{E}}$), which is mathematically equivalent to the conventional vacuum energy density ($p_{\textnormal{vac}}=-\rho_{\textnormal{vac}}$, where $\rho_{\textnormal{vac}}$ and $p_{\textnormal{vac}}$ are the energy density and pressure of the vacuum, respectively) for the spatially flat Robertson-Walker spacetime \cite{Sahni00}. We would like to note that the dynamics that would be generated by a positive cosmological constant in conventional $(1+3)$-dimensional general relativity is generated here by a negative cosmological constant we consider in the $(1+3+n)$-dimensional universe. We shall further comment on this below in Section 4.

We note that the universe has no beginning in the finite past and no end in the finite future and that neither the external nor internal dimensions reach zero size at finite $t$ values. Therefore, as the reference point to distinguish between different times, we consider the time $t_{\rm eq}=-\frac{3n}{(n+3)\tilde{n}\sqrt{\Lambda-k\rho_{0}}}\ln\left({\frac{a_{0}}{s_{0}}}\right)$ when the external and internal dimensions are at the same size $a(t_{\rm eq})=s(t_{\rm eq})$. We note that $a\sim s$, hence the space is effectively $(3+n)$-dimensional, when $t\sim t_{eq}$. On the other hand, provided that the internal space possesses a finite number of dimensions; $a\rightarrow 0$ and $s\rightarrow \infty$ as $t\rightarrow -\infty$, while $a\rightarrow \infty$ and $s\rightarrow 0$ as $t\rightarrow \infty$. Hence, although the $(3+n)$-dimensional space is eternal and preserves its constant volume, at very early times ($t<<t_{\rm eq}$) the space was effectively an infinitely large $n$-dimensional space, i.e., infinitely large $s$ and infinitely small $a$. Similarly, at very late times ($t>>t_{\rm eq}$) the space will be effectively an infinitely large $3$-dimensional space, i.e., infinitely large $a$ and infinitely small $s$.

From the constant $(3 + n)$-dimensional volume condition, it is evident that the magnitude of the expansion rate of the external space and of the contraction rate of the internal space are equal independent of the number of the internal dimensions. However, the number of the extra dimensions appears in two different ways in the Hubble parameters of the external and internal dimensions. For the external dimensions the factor is $\bar{n}$, which increases with the number of the extra dimensions, $\bar{n}=\sqrt{\frac{3}{2}},\sqrt{\frac{12}{5}},...,\sqrt{6}$ as $n=1,...,\infty$. For the internal dimensions, on the other hand, the factor is $\frac{\bar{n}}{n}$ which decreases with the number of extra dimensions, $\frac{\bar{n}}{n}=\sqrt{\frac{3}{2}},\sqrt{\frac{3}{5}},...,0$ as $n=1,...,\infty$. Accordingly, the number of the extra dimensions affects the dynamics of the external dimensions slightly, while it affects the dynamics of the internal dimensions drastically. The internal dimensions are almost static for very large values of $n$ and they even freeze as $n\rightarrow\infty$, although the external dimensions are affected only slightly. If we consider the limit $n\rightarrow\infty$, the extra dimensions will be static, for instance, at Planck length scales, but will still be affecting the evolution of the external dimensions. This is because, in that case, the total effect of the arbitrary number of extra dimensions with arbitrarirly small contractions have considerable effect on the dynamics of the external space. In other words, the scales of the extra dimensions could always be at Planck length scales, while the external dimensions expand monotonously to arbitrarily large scales.

\subsection{Solution for curved external and flat internal spaces ($\kappa_{a}\neq 0$, $\kappa_{s}=0$)}
\label{subsec:curvedflat}

In this section we give the solution for the case where the internal is space flat while the external space is curved. Accordingly, upon the substitution $\kappa_{s}=0$, the field equations (\ref{eqn:EFE1rr})-(\ref{eqn:EFE2rr}) read
\begin{equation}
\label{eqn:EFE1r2}
-\frac{3}{2}\left(\frac{3+n}{n}\right)\frac{\dot{a}^2}{a^2}+3\frac{\kappa_{a}}{a^2}+\Lambda=k\rho_{0},
\end{equation}
\begin{equation}
\label{eqn:EFE2r2}
\frac{1}{2}\left(\frac{9+5n}{n}\right)\frac{\dot{a}^2}{a^2}-\frac{\ddot{a}}{a}+\frac{\kappa_{a}}{a^2}+\Lambda=-kp .
\end{equation}
Solving (\ref{eqn:EFE1r2}), we obtain the cosmological parameters of the external dimensions as follows:
\begin{eqnarray}
\label{eqn:cfa}
a=a_{0}e^{\frac{\bar{n}}{3}\sqrt{\Lambda-k\rho_{0}} \; t}-\frac{3}{4}\frac{\kappa_{a}}{a_{0}(\Lambda-k\rho_{0})}e^{-\frac{\bar{n}}{3}\sqrt{\Lambda-k\rho_{0}} \; t},
\end{eqnarray}
\begin{eqnarray}
H_{a}=\frac{\bar{n}}{3}\sqrt{\Lambda-k\rho_{0}} \; \frac{a_{0} e^{2\frac{\bar{n}}{3}\sqrt{\Lambda-k\rho_{0}} \; t}+\frac{3}{4}\frac{\kappa_{a}}{a_{0}(\Lambda-k\rho_{0})}}{a_{0}e^{2\frac{\bar{n}}{3}\sqrt{\Lambda-k\rho_{0}} \; t}-\frac{3}{4}\frac{\kappa_{a}}{a_{0}(\Lambda-k\rho_{0})}},
\end{eqnarray}
\begin{eqnarray}
q_{a}=-\left(\frac{a_{0} e^{2\frac{\bar{n}}{3}\sqrt{\Lambda-k\rho_{0}} \; t}-\frac{3}{4}\frac{\kappa_{a}}{a_{0}(\Lambda-k\rho_{0})}}{a_{0} e^{2\frac{\bar{n}}{3}\sqrt{\Lambda-k\rho_{0}} \; t}+\frac{3}{4}\frac{\kappa_{a}}{a_{0}(\Lambda-k\rho_{0})}}\right)^2.
\end{eqnarray}
Using (\ref{eqn:cfa}) together with (\ref{eqn:ansatz1}), we obtain the cosmological parameters of the internal dimensions as follows:
\begin{eqnarray}
s=V_{0}^{\frac{1}{n}}\left(a_{0}e^{\frac{\bar{n}}{3}\sqrt{\Lambda-k\rho_{0}} \; t}-\frac{3}{4}\frac{\kappa_{a}}{a_{0}(\Lambda-k\rho_{0})}e^{-\frac{\bar{n}}{3}\sqrt{\Lambda-k\rho_{0}} \; t}\right)^{-\frac{3}{n}},
\end{eqnarray}
\begin{eqnarray}
H_{s}=-\frac{\bar{n}}{n}\sqrt{\Lambda-k\rho_{0}} \; \frac{a_{0} e^{2\frac{\bar{n}}{3}\sqrt{\Lambda-k\rho_{0}} \; t}+\frac{3}{4}\frac{\kappa_{a}}{a_{0}(\Lambda-k\rho_{0})}}{a_{0}e^{2\frac{\bar{n}}{3}\sqrt{\Lambda-k\rho_{0}} \; t}-\frac{3}{4}\frac{\kappa_{a}}{a_{0}(\Lambda-k\rho_{0})}},
\end{eqnarray}
\begin{eqnarray}
q_{s}=-1-\frac{n\frac{\kappa_{a}}{\Lambda-k\rho_{0}}e^{2\frac{\bar{n}}{3}\sqrt{\Lambda-k\rho_{0}} \; t}}{\left(a_{0} e^{2\frac{\bar{n}}{3}\sqrt{\Lambda-k\rho_{0}} \; t}+\frac{3}{4}\frac{\kappa_{a}}{a_{0}(\Lambda-k\rho_{0})}\right)^2}.
\end{eqnarray}
Finally, putting (\ref{eqn:cfa}) into (\ref{eqn:EFE2r2}), the pressure of the higher dimensional effective fluid is found to be:
\begin{eqnarray}
p&=&-\frac{\Lambda}{k}+\left(\rho_{0}-\frac{\Lambda}{k}\right)\left(\frac{a_{0} e^{2\frac{\bar{n}}{3}\sqrt{\Lambda-k\rho_{0}} \; t}+\frac{3}{4}\frac{\kappa_{a}}{a_{0}(\Lambda-k\rho_{0})}}{a_{0}e^{2\frac{\bar{n}}{3}\sqrt{\Lambda-k\rho_{0}} \; t}-\frac{3}{4}\frac{\kappa_{a}}{a_{0}(\Lambda-k\rho_{0})}} \right)^2
\nonumber\\
&&-\kappa_{a}\frac{3}{k}\frac{1+n}{3+n}{\left(a_{0}e^{\frac{\bar{n}}{3}\sqrt{\Lambda-k\rho_{0}} \; t}-\frac{3}{4}\frac{\kappa_{a}}{a_{0}(\Lambda-k\rho_{0})}e^{-\frac{\bar{n}}{3}\sqrt{\Lambda-k\rho_{0}} \; t}\right)^{-2}}.
\end{eqnarray}
Hence the EoS parameter of the higher dimensional effective fluid turns out to be
\begin{eqnarray}
w&=&-\frac{\Lambda}{k\rho_{0}}+\left(1-\frac{\Lambda}{k\rho_{0}}\right)\left(\frac{a_{0} e^{2\frac{\bar{n}}{3}\sqrt{\Lambda-k\rho_{0}} \; t}+\frac{3}{4}\frac{\kappa_{a}}{a_{0}(\Lambda-k\rho_{0})}}{a_{0}e^{2\frac{\bar{n}}{3}\sqrt{\Lambda-k\rho_{0}} \; t}-\frac{3}{4}\frac{\kappa_{a}}{a_{0}(\Lambda-k\rho_{0})}} \right)^2
\nonumber\\
&&-\kappa_{a}\frac{3}{k\rho_{0}}\frac{1+n}{3+n}{\left(a_{0}e^{\frac{\bar{n}}{3}\sqrt{\Lambda-k\rho_{0}} \; t}-\frac{3}{4}\frac{\kappa_{a}}{a_{0}(\Lambda-k\rho_{0})}e^{-\frac{\bar{n}}{3}\sqrt{\Lambda-k\rho_{0}} \; t}\right)^{-2}}.
\end{eqnarray}

One may observe that this solution is consistent for $\Lambda-k\rho_{0}>0$ in general, but is consistent for $\Lambda-k\rho_{0}<0$ only if $\kappa_{a}=1$ with a particular choice of the other parameters. The dynamics are quite different for the cases $\Lambda>k\rho_{0}$ and $\Lambda<k\rho_{0}$. Hence it will be convenient to discuss these two cases separately.

\begin{figure}[ht]

\begin{minipage}[b]{0.49\linewidth}
\psfrag{sfs}[b][b]{\footnotesize{$a$ and $s$}}
\psfrag{t}[b][b]{$t$}
\centering
\includegraphics[width=1\textwidth]{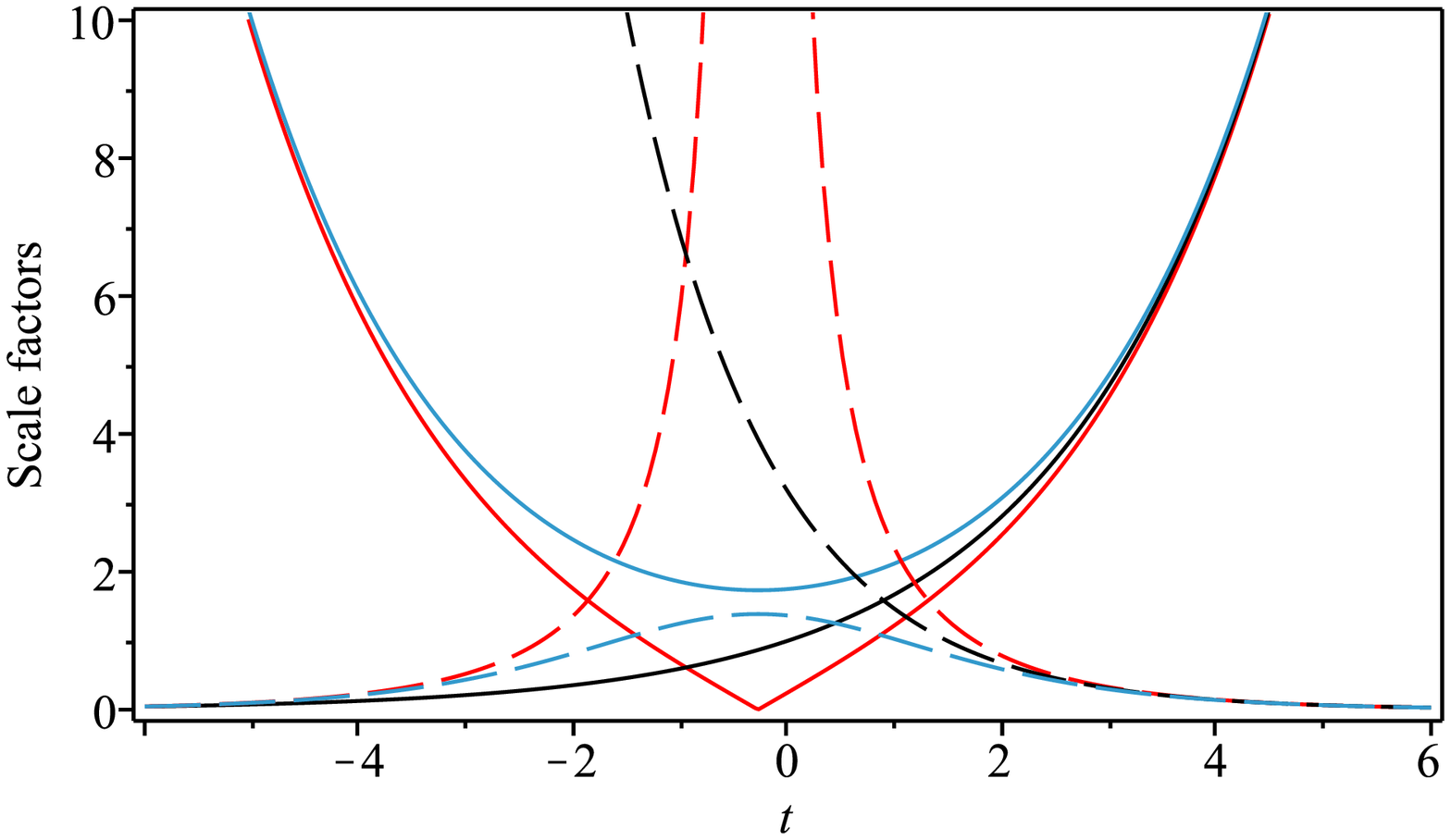}
\caption{Scale factors of the external and internal spaces versus cosmic time $t$ for $n=2$. Solid and dashed curves represent external and internal spaces respectively. Blue, black and red curves represent open, flat and closed external space cases respectively.}
\label{fig:alt}
\end{minipage}
\hspace{0.01\linewidth}
\begin{minipage}[b]{0.49\linewidth}
\centering
\includegraphics[width=1\textwidth]{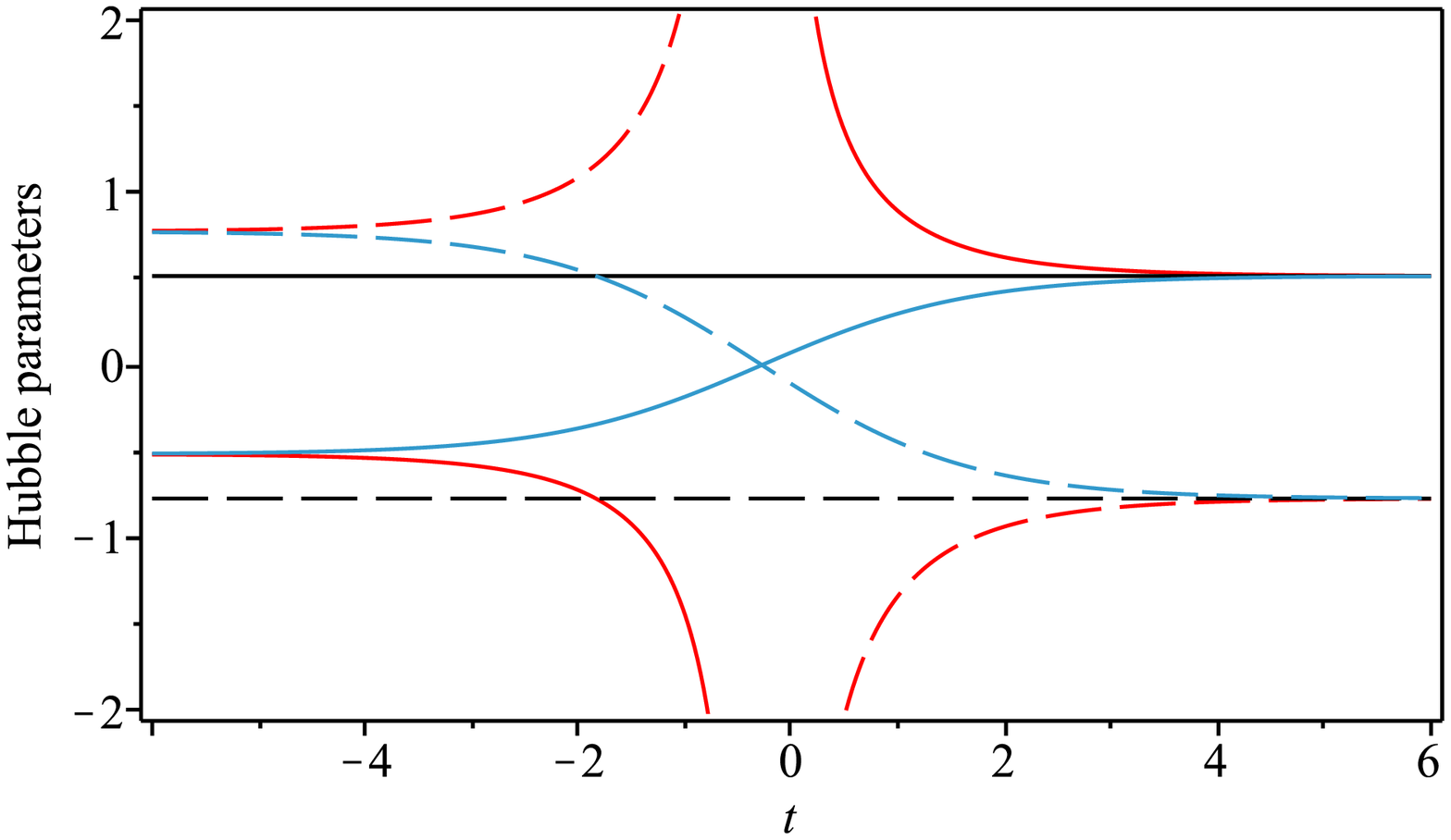}
\caption{Hubble parameters of the external and internal spaces versus cosmic time $t$ for $n=2$. Solid and dashed curves represent external and internal spaces respectively. Blue, black and red curves represent open, flat and closed external space cases respectively.}
\label{fig:halt}
\end{minipage}

\begin{minipage}[b]{0.49\linewidth}
\centering
\includegraphics[width=1\textwidth]{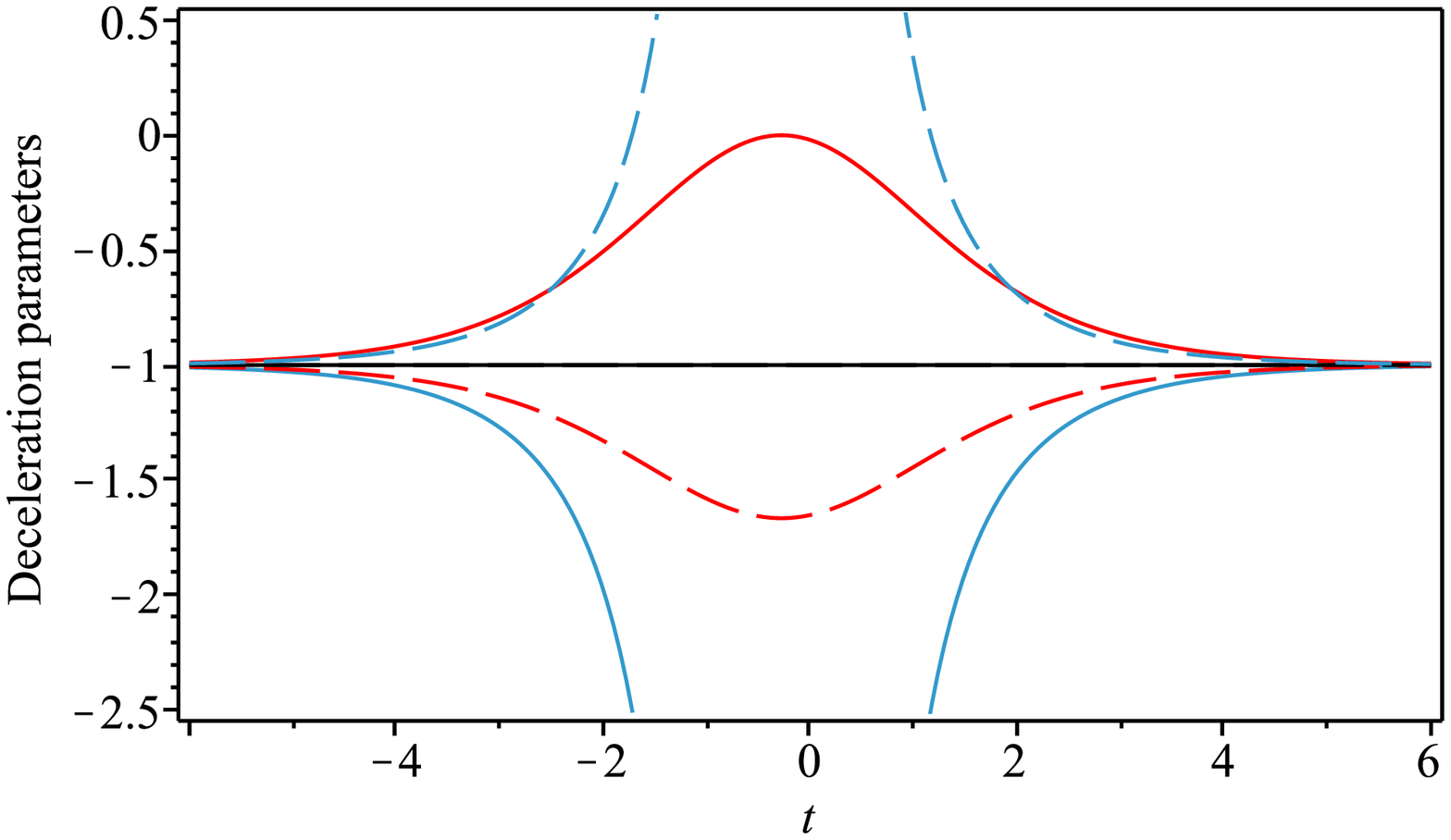}
\caption{Deceleration parameters of the external and internal spaces versus cosmic time $t$ for $n=2$. Solid and dashed curves represent external and internal spaces respectively. Blue, black and red curves represent open, flat and closed external space cases respectively.}
\label{fig:qalt}
\end{minipage}
\hspace{0.01\linewidth}
\begin{minipage}[b]{0.49\linewidth}
\centering
\includegraphics[width=1\textwidth]{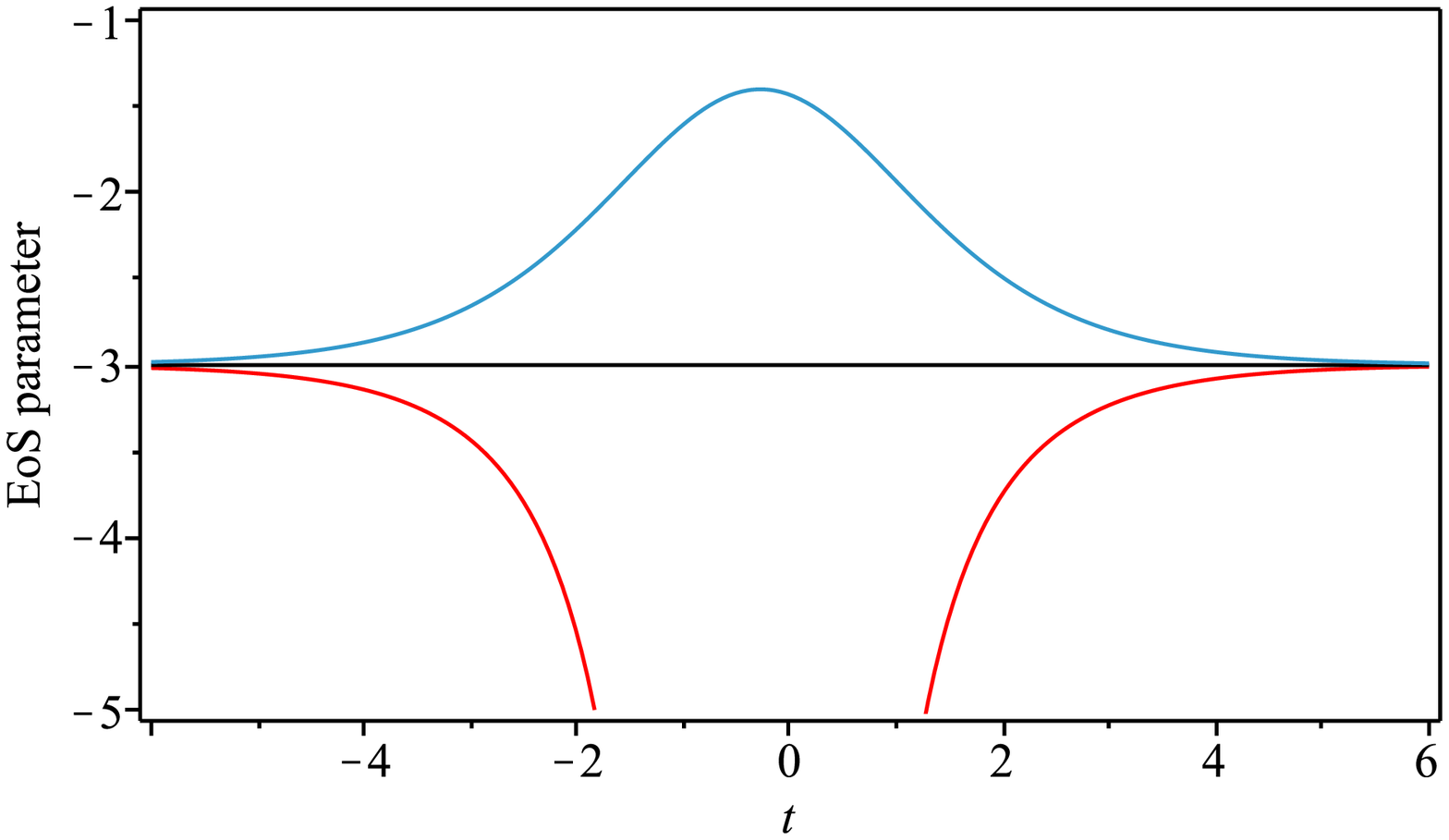}
\caption{EoS parameters of the external and internal spaces versus cosmic time $t$ for $n=2$. Solid and dashed curves represent external and internal spaces respectively. Blue, black and red curves represent open, flat and closed external space cases respectively.}
\label{fig:eose}
\end{minipage}

\end{figure}

First let us consider the case $\Lambda-k\rho_{0}>0$. We check that, for consistency, by substituting $\kappa_{a}=0$ the solution reduces to the one given for flat external and internal spaces. On the other hand, the non-flat geometry of the external space alters the exponential behavior of the solutions obtained for the case flat external and internal spaces. We find that there is a critical time
\begin{equation}
t_{{\rm{c}}}= \frac{3}{2\bar{n}}\frac{1}{\sqrt{\Lambda-\rho_{0}}}\ln\left(\frac{3}{4{a_{0}}^{3}\sqrt{\Lambda-k\rho_{0}}}\right),
\end{equation}
where
\begin{equation}
 \left. \begin{aligned}
a(t_{{\rm{c}}})&=0\\
\\
s(t_{{\rm{c}}})&=\infty
       \end{aligned}
 \right\}
 \quad\text{for }\kappa_{a}=1\text{ (closed external space)}
\end{equation}
and
\begin{equation}
 \left. \begin{aligned}
a(t_{{\rm{c}}})=a_{{\rm{min}}}&=\sqrt{\frac{3}{\Lambda-k\rho_{0}}}\\
\\
s(t_{{\rm{c}}})=s_{{\rm{max}}}&={V_{0}}^{\frac{1}{n}}\left(\frac{\Lambda-k\rho_{0}}{3}\right)^{\frac{3}{2n}}
       \end{aligned}
 \right\}
 \quad\text{for }\kappa_{a}=-1 \text{ (open external space)}.
\end{equation}
In both cases, the external space expands and the internal space contracts almost exponentially while $t\gg t_{{\rm{c}}}$. However, the non-zero curvature of the external space alters the dynamics dramatically for $t< t_{{\rm{c}}}$. So that the external space contracts and the internal space expands almost exponentially when $t\ll t_{{\rm{c}}}$. To show the generic behavior of the model, we depict the magnitudes of the scale factors (Fig. \ref{fig:alt}), Hubble parameters (Fig. \ref{fig:halt}) and deceleration parameters (Fig. \ref{fig:qalt}) of the external (solid lines) and internal (dashed lines) spaces. We have given the EoS parameter (Fig. \ref{fig:eose}) of the higher dimensional effective fluid for open (blue), flat (black) and closed (red) external spaces with the choice $n=2$ and some selected values for the parameters.

\textit{The case of open external space} ($\kappa_{a}=-1$): The negative curvature of the external space prevents it from ever reaching zero size (i.e., $a_{\rm min}>0$). In the infinite past, the external space is infinitely large and the internal space is infinitely small. The external dimensions bounce\footnote{See Ref. \cite{Novello08} and references therein for further reading on bouncing cosmologies.} at $t=t_{{\rm{c}}}$; they contract till they reach a minimum value $a_{{\rm{min}}}$ at $t=t_{{\rm{c}}}$ and then start expanding at a rate that approximates the de Sitter expansion for $t\gg t_{{\rm{c}}}$. Meanwhile the internal dimensions expand till they reach a maximum value $s_{{\rm{max}}}$ at $t=t_{{\rm{c}}}$ and then start contracting at a rate that approximates exponential contraction. In this solution we have two interesting observations to make. Expansion of the external dimensions starts from a non-zero size $a_{\rm min}$, which may be called the non-singular Big Bang for an observer living today in an effectively $(1+3)$-dimensional universe. Following this early expansion, the external dimensions enter into a de Sitter expansion, which may be related with inflation in the early universe. On the other hand, the size of the internal dimensions can never exceed a maximum value $s_{{\rm{max}}}$. Hence, although the internal dimensions affect the dynamics of the external dimensions, they may never be able to reach observable scales.

\textit{The case of closed external space} ($\kappa_{a}=1$): The dynamics in this solution are similar to the ones we obtain in open external space case when $t\nsim t_{\rm c}$, but are different for $t\sim t_{\rm c}$; the positive curvature of the external space forces it to reach zero size (i.e., $a_{\rm min}=0$) at $t_{\rm c}$. In the infinite past, the external dimensions are infinitely large and the internal dimensions are infinitely small. The external dimensions contract and reach zero size at $t=t_{{\rm{c}}}$, while the internal dimensions expand and become infinitely large. The expansion of the external dimensions starts from zero size at $t=t_{{\rm{c}}}$ and goes at a rate that approximates the de Sitter expansion for $t\gg t_{{\rm{c}}}$, while the contraction of the internal dimensions starts from an infinitely large size and goes at a rate that approximates exponential contraction. The external dimensions start expanding from zero size at $t=t_{{\rm{c}}}$ with $q_{a}=0$ and $H_{a}=\infty$, i.e. there is a Big Bang at $t=t_{{\rm{c}}}$ for the external space and that can be taken as the origin of the universe by an observer living today in an effectively $(1+3)$-dimensional universe. Following the Big Bang, the external dimensions evolve into de Sitter expansion, i.e. the inflationary regime as $t$ increases.

Similar to the solution with both the external and internal dimensions flat, the number of internal dimensions here alters the dynamics of the external space only slightly but the dynamics of the internal space drastically. In the limit $n\rightarrow\infty$, while the external dimensions preserve the generic pattern of their expansion, the internal dimensions freeze at unit size, i.e., $s\rightarrow 1$ and $H_{s}\rightarrow 0$.

In the case of $\Lambda-k\rho_{0}<0$, using the Euler's formula, the scale factor of the external space can be written as follows:
\begin{eqnarray}
a=a_{0}\left(1+\frac{3}{4}\frac{\kappa_{a}}{{a_{0}}^{2}(k\rho_{0}-\Lambda)}\right)\cos\left(\frac{\bar{n}}{3}\sqrt{k\rho_{0}-\Lambda} \; t\right)+i a_{0} \left(1-\frac{3}{4}\frac{\kappa_{a}}{{a_{0}}^{2}(k\rho_{0}-\Lambda)}\right)\sin\left(\frac{\bar{n}}{3}\sqrt{k\rho_{0}-\Lambda} \; t\right),
\end{eqnarray}
which is a complex function. However, provided that
\begin{equation}
\kappa_{a}=1\quad\text{and}\quad a_{0}=\frac{1}{2}\sqrt{\frac{3}{k\rho_{0}-\Lambda}},
\end{equation}
the imaginary part vanishes and we obtain a real solution. Hence, using these values for the parameters, we obtain the cosmological parameters of the external dimensions as follows:
\begin{equation}
a=\sqrt{\frac{3}{k\rho_{0}-\Lambda}}\cos\left(\frac{\bar{n}}{3}\sqrt{k\rho_{0}-\Lambda} \; t\right),
\end{equation}
\begin{equation}
H_{a}=-\frac{\bar{n}}{3}\sqrt{k\rho_{0}-\Lambda}\tan\left(\frac{\bar{n}}{3}\sqrt{k\rho_{0}-\Lambda} \; t\right),
\end{equation}
\begin{equation}
q_{a}=-1+\sin^{-2}\left(\frac{\bar{n}}{3}\sqrt{k\rho_{0}-\Lambda} \; t\right),
\end{equation}
and of the internal dimensions as follows:
\begin{equation}
s={V_{0}}^{\frac{1}{n}}\left(\frac{3}{k\rho_{0}-\Lambda}\right)^{-\frac{3}{2n}}\cos^{-\frac{3}{n}}\left(\frac{\bar{n}}{3}\sqrt{k\rho_{0}-\Lambda} \; t\right),
\end{equation}
\begin{equation}
H_{s}=\frac{\bar{n}}{n}\sqrt{k\rho_{0}-\Lambda}\tan\left(\frac{\bar{n}}{3}\sqrt{k\rho_{0}-\Lambda} \; t\right),
\end{equation}
\begin{equation}
q_{s}=-1-\frac{n}{3}\sin^{-2}\left(\frac{\bar{n}}{3}\sqrt{k\rho_{0}-\Lambda} \; t\right).
\end{equation}
The pressure and EoS parameter of the higher dimensional effective fluid become:
\begin{equation}
p=\rho_{0}-2\frac{\Lambda}{k}-\left(\rho_{0}-\frac{\Lambda}{k}\right)\frac{4+2n}{3+n}\cos^{-2}\left(\frac{\bar{n}}{3}\sqrt{k\rho_{0}-\Lambda} \; t\right),
\end{equation}
\begin{equation}
w=1-2\frac{\Lambda}{k\rho_{0}}-\left(1-\frac{\Lambda}{k\rho_{0}}\right)\frac{4+2n}{3+n}\cos^{-2}\left(\frac{\bar{n}}{3}\sqrt{k\rho_{0}-\Lambda} \; t\right).
\end{equation}

This oscillating\footnote{See Ref. \cite{Novello08,Barrow95OS,Sahni12} and references therein for further reading on oscillating cosmologies.} special case is possible if the external space is closed but the internal space is flat. Since $\Lambda<k\rho_{0}$, the $\Lambda$ can be less than zero, which corresponds to a $(1+3+n)$-dimensional positive cosmological constant in this case. The external and internal dimensions oscillate with a period $T=\frac{6\pi}{\bar{n}\sqrt{k\rho_{0}-\Lambda}}$ and with amplitudes $A_{a}=\sqrt{\frac{3}{k\rho_{0}-\Lambda}}$ and $A_{s}={V_{0}}^{\frac{1}{n}}\left(\frac{k\rho_{0}-\Lambda}{3}\right)^{\frac{3}{2n}}$, respectively. The number of the internal dimensions does not alter the amplitude of the external dimensions at all but the amplitude of the internal dimensions are affected strongly. It also alters the period of the oscillations slightly. In the limit $n\rightarrow\infty$; the period of oscillations $T\rightarrow\frac{\pi}{\sqrt{6}}\frac{1}{\sqrt{k\rho_{0}-\Lambda}}$ and the internal dimensions oscillate with unit amplitude $A_{s}\rightarrow 1$.

\subsection{Solution for flat external and curved internal spaces ($\kappa_{a}= 0$, $\kappa_{s}\neq 0$)}
\label{subsec:flatcurved}

In this section we discuss the solutions where the external space is flat while the internal space is curved and $n\geq2$ (We note that for $n=1$ the solution in this section reduces to the solution for which both the external and internal spaces are flat given in section \ref{subsec:flatflat}). Accordingly, substituting $\kappa_{a}=0$, the field equations (\ref{eqn:EFE1rr})-(\ref{eqn:EFE2rr}) read
\begin{equation}
\label{eqn:EFE1r3}
-\frac{3}{2}\left(\frac{3+n}{n}\right)\frac{\dot{a}^2}{a^2}+\frac{1}{2}\kappa_{s} n(n-1)V_{0}^{-\frac{2}{n}}a^{\frac{6}{n}}+\Lambda=k\rho_{0},
\end{equation}
\begin{equation}
\label{eqn:EFE2r3}
\frac{1}{2}\left(\frac{9+5n}{n}\right)\frac{\dot{a}^2}{a^2}-\frac{\ddot{a}}{a}+\frac{1}{2}\kappa_{s} n(n-1) V_{0}^{-\frac{2}{n}}a^{\frac{6}{n}}+\Lambda=-kp .
\end{equation}
Solving (\ref{eqn:EFE1r3}), the cosmological parameters of the external dimensions are obtained as follows:
\begin{equation}
\label{eqn:afle}
a=V_{0}^{\frac{1}{3}}\left(s_{0}e^{-\frac{\bar{n}}{n}\sqrt{\Lambda-k\rho_{0}} \; t}-\frac{n(n-1)}{8}\frac{\kappa_{s}}{s_{0}(\Lambda-k\rho_{0})}e^{\frac{\bar{n}}{n}\sqrt{\Lambda-k\rho_{0}} \; t}\right)^{-\frac{n}{3}},
\end{equation}
\begin{equation}
H_{a}=\frac{\bar{n}}{3}\sqrt{\Lambda-k\rho_{0}}\frac{s_{0}e^{-2\frac{\bar{n}}{n}\sqrt{\Lambda-k\rho_{0}} \; t}+\frac{n(n-1)}{8}\frac{\kappa_{s}}{s_{0}(\Lambda-k\rho_{0})}}{s_{0}e^{-2\frac{\bar{n}}{n}\sqrt{\Lambda-k\rho_{0}} \; t}-\frac{n(n-1)}{8}\frac{\kappa_{s}}{s_{0}(\Lambda-k\rho_{0})}},
\end{equation}
\begin{eqnarray}
q_{a}=-1-\frac{(n-1)\frac{6}{4}\frac{\kappa_{s}}{\Lambda-k\rho_{0}}e^{-2\frac{\bar{n}}{n}\sqrt{\Lambda-k\rho_{0}} \; t}}{\left(s_{0}e^{-2\frac{\bar{n}}{n}\sqrt{\Lambda-k\rho_{0}} \; t}+\frac{n(n-1)}{8}\frac{\kappa_{s}}{s_{0}(\Lambda-k\rho_{0})}\right)^2}.
\end{eqnarray}
Using (\ref{eqn:afle}) together with (\ref{eqn:ansatz1}), we obtain the cosmological parameters of the internal dimensions as follows:
\begin{equation}
\label{eqn:sfle}
s=s_{0}e^{-\frac{\bar{n}}{n}\sqrt{\Lambda-k\rho_{0}} \; t}-\frac{n(n-1)}{8}\frac{\kappa_{s}}{s_{0}(\Lambda-k\rho_{0})}e^{\frac{\bar{n}}{n}\sqrt{\Lambda-k\rho_{0}} \; t},
\end{equation}
\begin{equation}
H_{s}=-\frac{\bar{n}}{n}\sqrt{\Lambda-k\rho_{0}}\frac{s_{0}e^{-2\frac{\bar{n}}{n}\sqrt{\Lambda-k\rho_{0}} \; t}+\frac{n(n-1)}{8}\frac{\kappa_{s}}{s_{0}(\Lambda-k\rho_{0})}}{s_{0}e^{-2\frac{\bar{n}}{n}\sqrt{\Lambda-k\rho_{0}} \; t}-\frac{n(n-1)}{8}\frac{\kappa_{s}}{s_{0}(\Lambda-k\rho_{0})}},
\end{equation}
\begin{equation}
q_{s}=-\left(\frac{s_{0}e^{-2\frac{\bar{n}}{n}\sqrt{\Lambda-k\rho_{0}} \; t}-\frac{n(n-1)}{8}\frac{\kappa_{s}}{s_{0}(\Lambda-k\rho_{0})}}{s_{0}e^{-2\frac{\bar{n}}{n}\sqrt{\Lambda-k\rho_{0}} \; t}+\frac{n(n-1)}{8}\frac{\kappa_{s}}{s_{0}(\Lambda-k\rho_{0})}}\right)^2.
\end{equation}
Finally, using $a$ in (\ref{eqn:EFE2r2}), the pressure of the higher dimensional effective fluid is found to be:
\begin{eqnarray}
\label{eqn:pressurefle}
p&=&-\frac{\Lambda}{k}+\left(\rho_{0}-\frac{\Lambda}{k} \right)\left(\frac{s_{0}e^{-2\frac{\bar{n}}{n}\sqrt{\Lambda-k\rho_{0}} \; t}+\frac{n(n-1)}{8}\frac{\kappa_{s}}{s_{0}(\Lambda-k\rho_{0})}}{s_{0}e^{-2\frac{\bar{n}}{n}\sqrt{\Lambda-k\rho_{0}} \; t}-\frac{n(n-1)}{8}\frac{\kappa_{s}}{s_{0}(\Lambda-k\rho_{0})}} \right)^{2}
\nonumber\\
&&-\kappa_{s}\frac{3}{k}\frac{1+n}{3+n}V_{0}^{-\frac{2}{3}}\left(s_{0}e^{-\frac{\bar{n}}{n}\sqrt{\Lambda-k\rho_{0}} \; t}-\frac{n(n-1)}{8}\frac{\kappa_{s}}{s_{0}(\Lambda-k\rho_{0})}e^{\frac{\bar{n}}{n}\sqrt{\Lambda-k\rho_{0}} \; t}\right)^{\frac{2}{3}n}.
\end{eqnarray}
Hence the EoS parameter of the fluid is also found:
\begin{eqnarray}
w&=&-\frac{\Lambda}{k\rho_{0}}+\left(1-\frac{\Lambda}{k\rho_{0}} \right)\left(\frac{s_{0}e^{-2\frac{\bar{n}}{n}\sqrt{\Lambda-k\rho_{0}} \; t}+\frac{n(n-1)}{8}\frac{\kappa_{s}}{s_{0}(\Lambda-k\rho_{0})}}{s_{0}e^{-2\frac{\bar{n}}{n}\sqrt{\Lambda-k\rho_{0}} \; t}-\frac{n(n-1)}{8}\frac{\kappa_{s}}{s_{0}(\Lambda-k\rho_{0})}} \right)^{2}
\nonumber\\
&&-\kappa_{s}\frac{3}{k\rho_{0}}\frac{1+n}{3+n}V_{0}^{-\frac{2}{3}}\left(s_{0}e^{-\frac{\bar{n}}{n}\sqrt{\Lambda-k\rho_{0}} \; t}-\frac{n(n-1)}{8}\frac{\kappa_{s}}{s_{0}(\Lambda-k\rho_{0})}e^{\frac{\bar{n}}{n}\sqrt{\Lambda-k\rho_{0}} \; t}\right)^{\frac{2}{3}n}.
\end{eqnarray}

Similar to the solution for the curved external and flat internal spaces, this solution is also consistent for $\Lambda-k\rho_{0}>0$ in general, but is consistent for $\Lambda-k\rho_{0}<0$ only if $\kappa_{s}=1$ with a particular choice of the other parameters. In this solution also, the dynamics are quite different for the cases $\Lambda>k\rho_{0}$ and $\Lambda<k\rho_{0}$. Hence it will be convenient to discuss the physical behavior of the model in these two cases separately.

\begin{figure}[ht]

\begin{minipage}[b]{0.49\linewidth}
\centering
\includegraphics[width=1\textwidth]{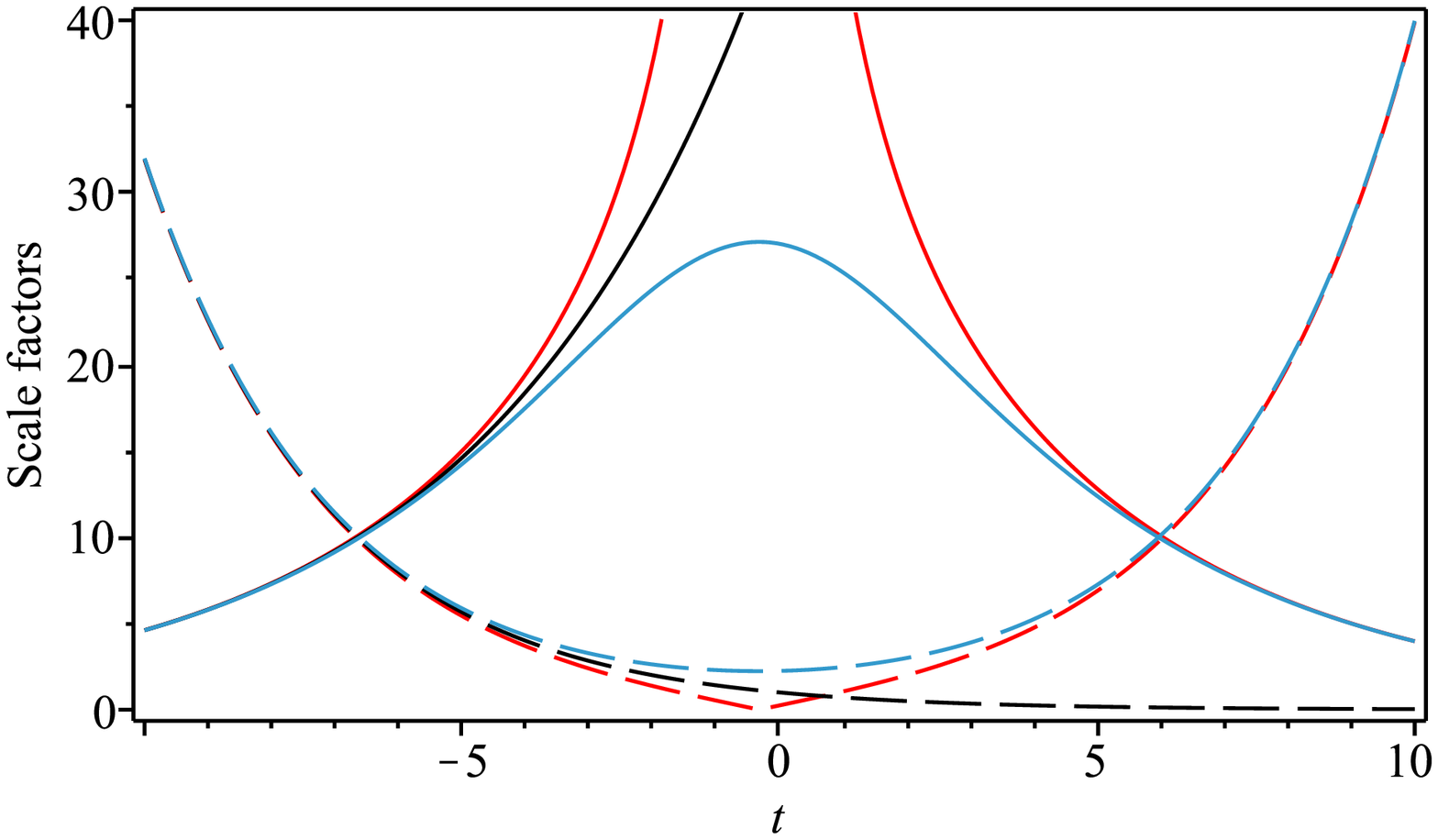}
\caption{Scale factors of the external and internal spaces versus cosmic time $t$ for $n=2$. Solid and dashed curves represent external and internal spaces respectively. Blue, black and red curves represent open, flat and closed internal space cases respectively.}
\label{fig:aet}
\end{minipage}
\hspace{0.01\linewidth}
\begin{minipage}[b]{0.49\linewidth}
\centering
\includegraphics[width=1\textwidth]{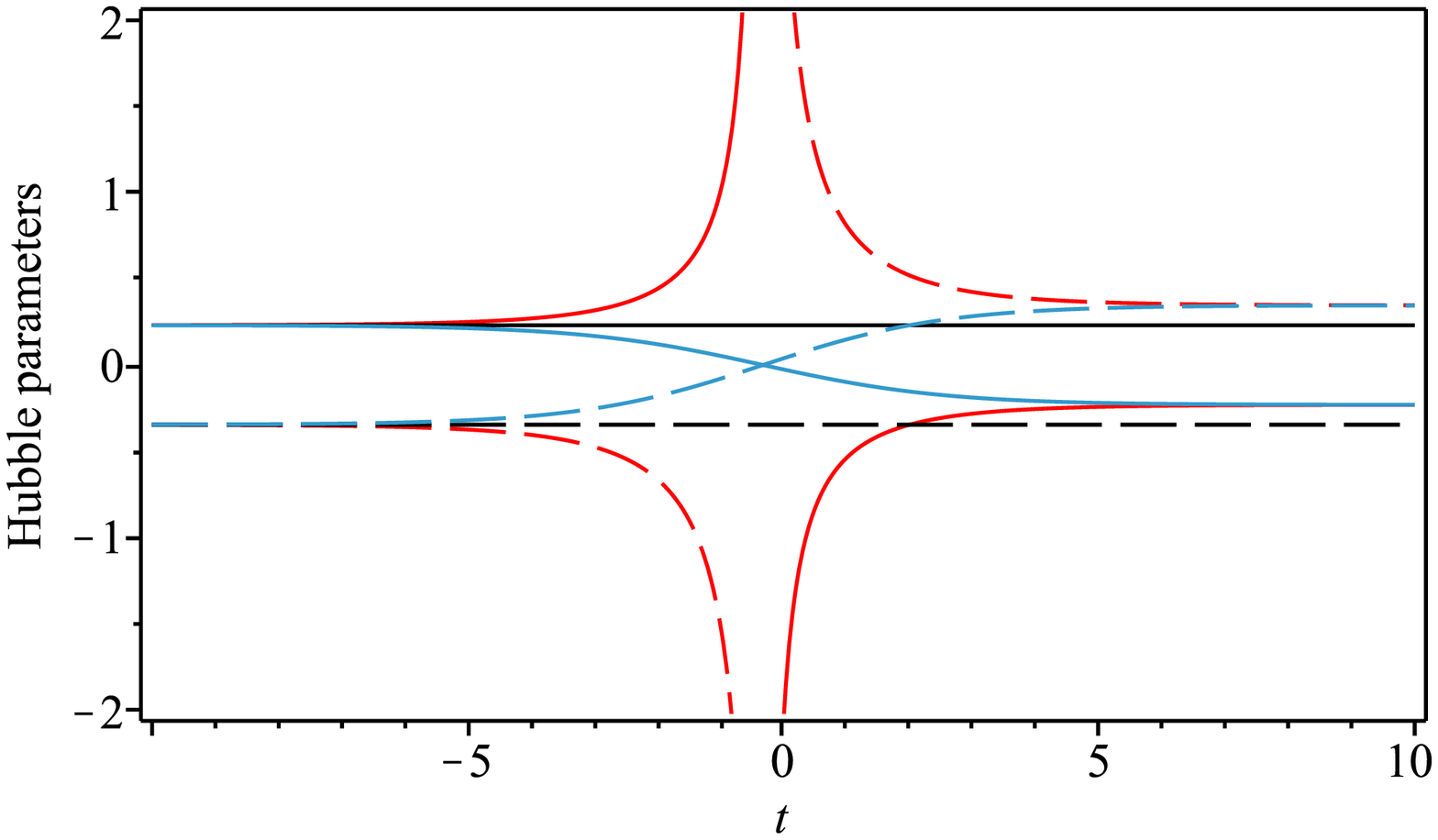}
\caption{Hubble parameters of the external and internal spaces versus cosmic time $t$ for $n=2$. Solid and dashed curves represent external and internal spaces respectively. Blue, black and red curves represent open, flat and closed internal space cases respectively.}
\label{fig:haet}
\end{minipage}

\begin{minipage}[b]{0.49\linewidth}
\centering
\includegraphics[width=1\textwidth]{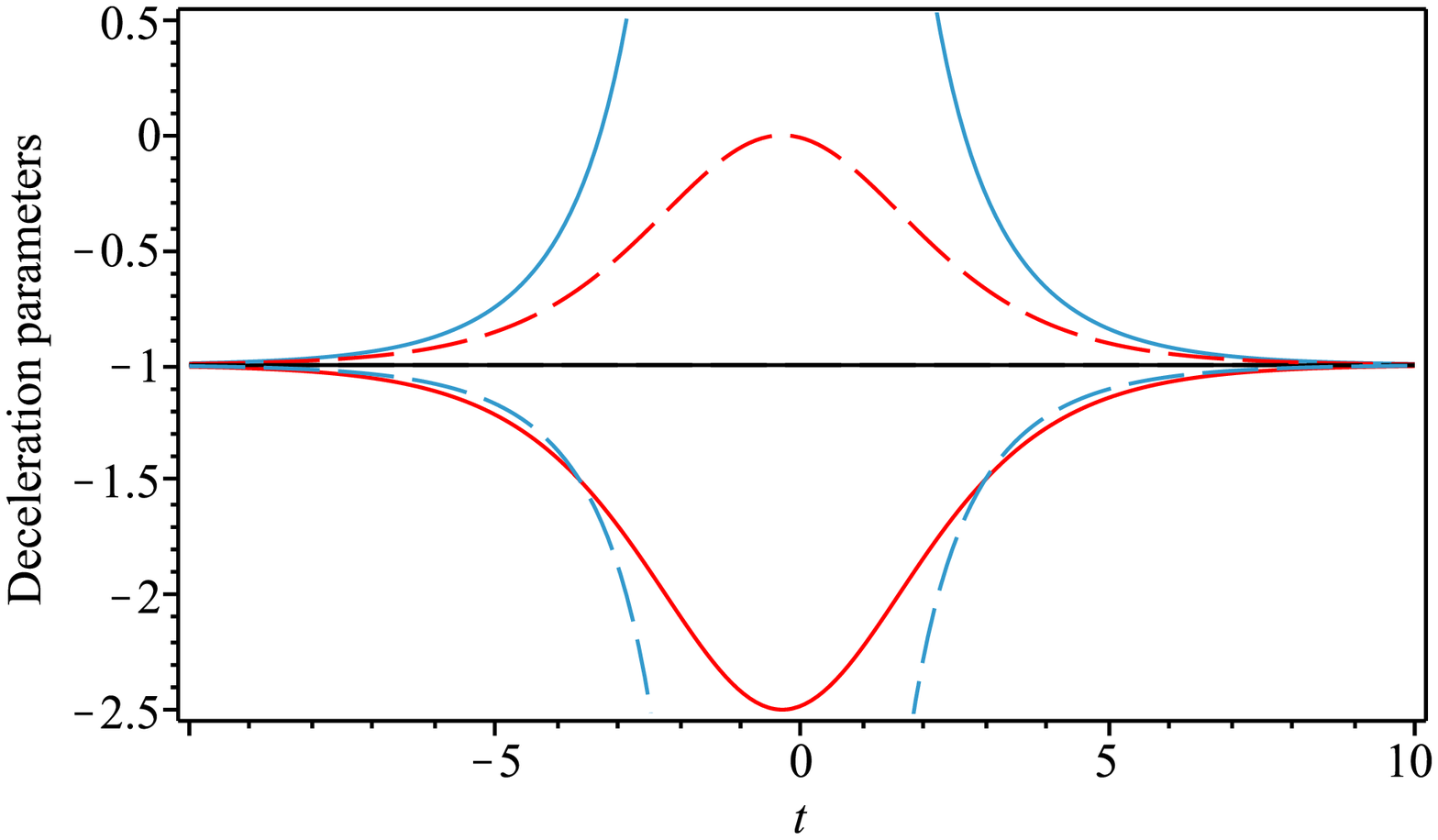}
\caption{Deceleration parameters of the external and internal spaces versus cosmic time $t$ for $n=2$. Solid and dashed curves represent external and internal spaces respectively. Blue, black and red curves represent open, flat and closed internal space cases respectively.}
\label{fig:qaet}
\end{minipage}
\hspace{0.01\linewidth}
\begin{minipage}[b]{0.49\linewidth}
\centering
\includegraphics[width=1\textwidth]{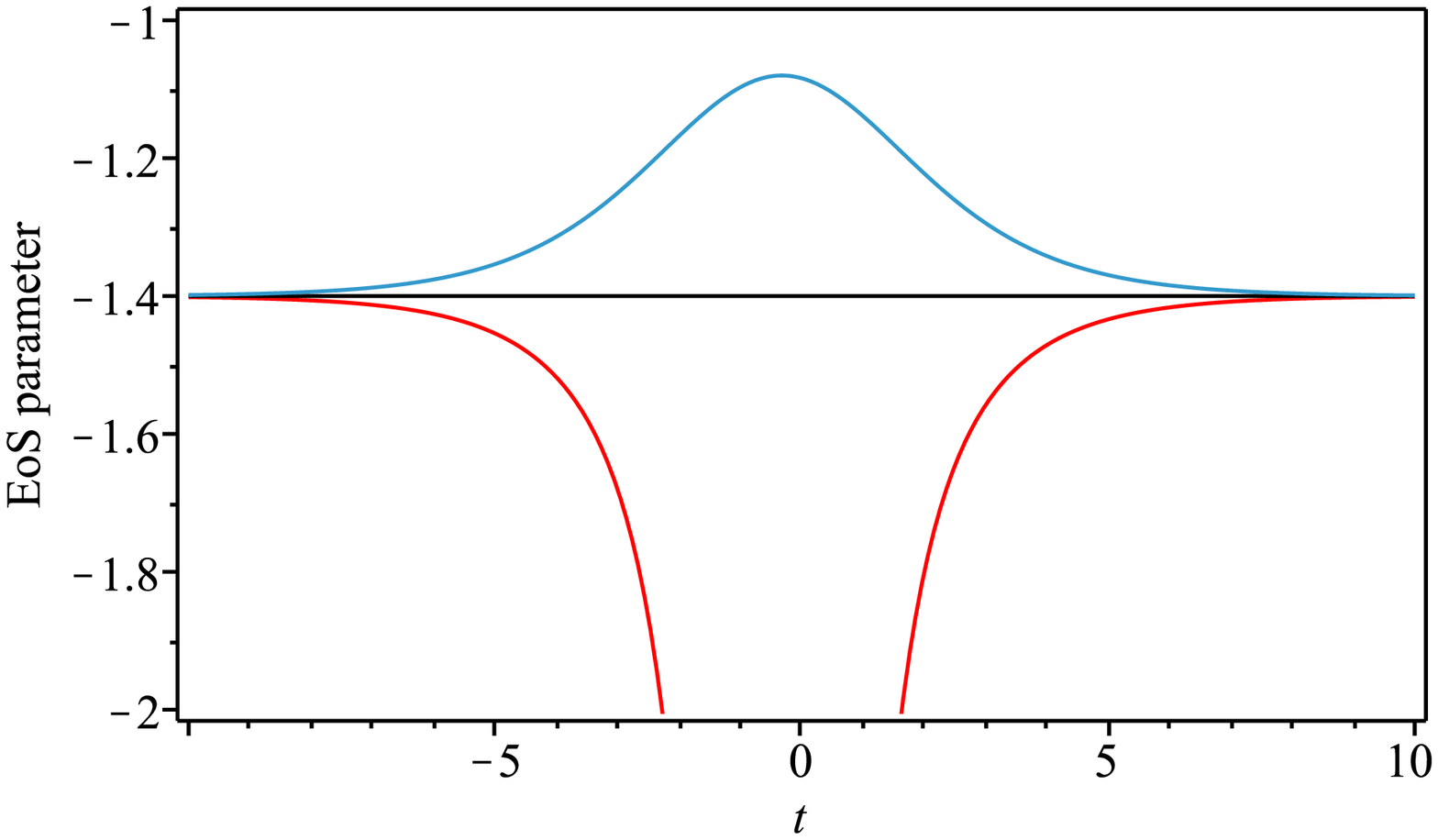}
\caption{EoS parameters of the external and internal spaces versus cosmic time $t$ for $n=2$. Solid and dashed curves represent external and internal spaces respectively. Blue, black and red curves represent open, flat and closed internal space cases respectively.}
\label{fig:eosl}
\end{minipage}

\end{figure}

First considering the case $\Lambda-k\rho_{0}>0$, one may check, for consistency, by substituting $\kappa_{s}=0$ the solution reduces as expected to the one given in \cite{DereliTucker83} when both the external and internal spaces are flat. The exponential behavior in the solution for flat external and internal spaces are altered, this time, by the non-flat geometry of the internal space. In this solution too, there is a critical time
\begin{eqnarray}
t_{{\rm{c}}}=\sqrt{\frac{2}{\bar{n}}}\frac{1}{\sqrt{\Lambda-k\rho_{0}}}\ln\left(8{s_{0}}^2 \frac{\Lambda-k\rho_{0}}{n(n-1)} \right) 
\end{eqnarray}
where
\begin{equation}
 \left. \begin{aligned}
a(t_{{\rm{c}}})&=\infty\\
\\
s(t_{{\rm{c}}})&=0
       \end{aligned}
 \right\}
  \quad\text{for }\kappa_{s}=1\text{ (closed internal space)}
\end{equation}
and
\begin{equation}
 \left. \begin{aligned}
a(t_{{\rm{c}}})=a_{{\rm{max}}}&=V_{0}^{\frac{1}{3}}\left(2\frac{\Lambda-k\rho_{0}}{n(n-1)}\right)^{\frac{n}{6}}\\
\\
s(t_{{\rm{c}}})=s_{{\rm{min}}}&=\sqrt{\frac{1}{2}\frac{n(n-1)}{\Lambda-k\rho_{0}}}
       \end{aligned}
 \right\}
  \quad\text{for }\kappa_{s}=-1\text{ (open internal space)}.
\end{equation}
In both cases of closed and open internal spaces, when $t\ll t_{{\rm{c}}}$ the external dimensions expand almost exponentially while the internal dimensions contract almost exponentially. However, as would be expected, the curvature of the internal space becomes more dominant as the internal space contracts and eventually develops a drastic deviation from the exponential behavior. If the internal space is closed, it reaches zero size ($s_{\rm min}=0$) at $t=t_{{\rm{c}}}$ while the external space reaches infinitely large sizes. If the internal space is open, on the other hand, it reaches its non-zero minimum value $s_{{\rm{min}}}$ while the external space expands until it reaches its finite maximum value $a_{{\rm{max}}}$ at $t=t_{{\rm{c}}}$ and then starts contracting. To show the generic behavior of the model, we depict the magnitudes of the scale factors (Fig. \ref{fig:aet}), Hubble parameters (Fig. \ref{fig:haet}) and deceleration parameters (Fig. \ref{fig:qaet}) of the external (solid curves) and internal (dashed curves) dimensions. We also give the EoS parameter (Fig. \ref{fig:eosl}) of the higher dimensional effective fluid for the case of open (blue), flat (black) and closed (red) internal spaces with the choice $n=2$ and some selected values for the parameters.

In this solution, the size of the internal dimensions do not remain small. Moreover, in contrast with the solutions in the previous sections, the number of the internal dimensions suppress the expansion of the external dimensions and the contraction of the internal dimensions. As $n\rightarrow\infty$; $s \rightarrow \infty$ and $H_{s}\rightarrow +\infty$, whereas $a\rightarrow 0$ and $H_{a}\rightarrow -\infty$. This behavior is understandable; the spatial curvature of the internal space $\kappa_{s}n(n-1)/s^2$ becomes infinitely large unless $s$ becomes infinitely large too. Hence a curved internal space with arbitrarily large number of dimensions is fatal for a dynamical cosmology.

In the case of $\Lambda-k\rho_{0}<0$, using the Euler's formula, the scale factor of the internal space can be written as follows:
\begin{eqnarray}
s=s_{0}\left(1+\frac{n(n-1)}{8}\frac{\kappa_{s}}{{s_{0}}^2(k\rho_{0}-\Lambda)}\right)\cos\left(\frac{\bar{n}}{n}\sqrt{k\rho_{0}-\Lambda}\;t\right)+is_{0}\left(1-\frac{n(n-1)}{8}\frac{\kappa_{s}}{{s_{0}}^2(k\rho_{0}-\Lambda)}\right)\sin\left(\frac{\bar{n}}{n}\sqrt{k\rho_{0}-\Lambda}\;t\right)
\end{eqnarray}
which is a complex function. Similar to the discussion in section \ref{subsec:curvedflat}, for the choice
\begin{equation}
n\geq 2,\quad\kappa_{s}=1\quad\text{and}\quad s_{0}=\sqrt{\frac{n(n-1)}{8(k\rho_{0}-\Lambda)}},
\end{equation}
the imaginary part vanishes and a real solution is obtained. Hence, using these values for the parameters, we obtain the cosmological parameters of the external dimensions as:
\begin{equation}
a={V_{0}}^{\frac{1}{3}}\left(\frac{1}{2}\frac{n(n-1)}{k\rho_{0}-\Lambda}\right)^{-\frac{n}{6}}\cos^{-\frac{n}{3}}\left(\frac{\bar{n}}{n}\sqrt{k\rho_{0}-\Lambda}\;t \right),
\end{equation}
\begin{equation}
H_{a}=\frac{\bar{n}}{3}\sqrt{k\rho_{0}-\Lambda}\tan\left(\frac{\bar{n}}{n}\sqrt{k\rho_{0}-\Lambda}\;t \right),
\end{equation}
\begin{equation}
q_{a}=-1-\frac{3}{n}\sin^{-2}\left(\frac{\bar{n}}{n}\sqrt{k\rho_{0}-\Lambda}\;t \right),
\end{equation}
the cosmological parameters of the internal dimensions as:
\begin{equation}
s=\sqrt{\frac{1}{2}\frac{n(n-1)}{k\rho_{0}-\Lambda}}\cos\left(\frac{\bar{n}}{n}\sqrt{k\rho_{0}-\Lambda}\;t \right),
\end{equation}
\begin{equation}
H_{s}=-\frac{\bar{n}}{n}\sqrt{k\rho_{0}-\Lambda}\tan\left(\frac{\bar{n}}{n}\sqrt{k\rho_{0}-\Lambda}\;t \right),
\end{equation}
\begin{equation}
q_{s}=-1+\sin^{-2}\left(\frac{\bar{n}}{n}\sqrt{k\rho_{0}-\Lambda}\;t \right),
\end{equation}
and the pressure and EoS parameter of the higher dimensional effective fluid as:
\begin{equation}
p=\rho_{0}-2\frac{\Lambda}{k}-\left(\rho_{0}-\frac{\Lambda}{k}\right)\frac{4+2n}{3+n}\cos^{-2}\left(\frac{\bar{n}}{n}\sqrt{k\rho_{0}-\Lambda}\;t \right),
\end{equation}
\begin{equation}
w=1-2\frac{\Lambda}{k\rho_{0}}-\left(1-\frac{\Lambda}{k\rho_{0}}\right)\frac{4+2n}{3+n}\cos^{-2}\left(\frac{\bar{n}}{n}\sqrt{k\rho_{0}-\Lambda}\;t \right).
\end{equation}

This oscillating special case is possible provided that the internal space is closed but the external space is flat. Since $\Lambda<k\rho_{0}$, the $\Lambda$ can be less than zero, which corresponds to a $(1+3+n)$-dimensional positive cosmological constant in this case. The external and internal dimensions oscillate continuously with the amplitudes $A_{a}={V_{0}}^{\frac{1}{3}}\left(\frac{1}{2}\frac{n(n-1)}{k\rho_{0}-\Lambda}\right)^{-\frac{n}{6}}$ and $A_{s}=\sqrt{\frac{1}{2}\frac{n(n-1)}{k\rho_{0}-\Lambda}}$, respectively, and with a period $T=\frac{2\pi n}{\bar{n}\sqrt{k\rho_{0}-\Lambda}}$. In this solution, contrary to the case of oscillating solution for the curved external and flat internal spaces; the higher the number of internal dimensions the longer the period of oscillations. As $n\rightarrow\infty$; the period of oscillations $T\rightarrow \infty$, which means that there will no oscillation anymore. In the same limits the amplitude of the external space $A_{a}\rightarrow 0$ and $H_{a}\rightarrow 0$, which means that the external dimensions freeze at zero size, while $A_{s}\rightarrow\infty$ with $H_{s}\rightarrow 0$, which means that the internal dimensions freeze at infinitely large size similar to the non-oscillating solution above.

\section{The universe according to an observer in $(1+3)$-dimensions}
\label{observer}

We do not usually deal in cosmology with direct measurements of
energy density and pressure of the material/physical content of the
universe. We collect data on the kinematics of the observed universe
instead, for instance from the supernova Ia observations
\cite{Perlmutter99} and on the geometry of the
space from cosmic microwave background by WMAP observations
\cite{Komatsu11}. Furthermore, we assume that the space we live in
is three dimensional. Then, what we do in general is
to interpret the collected information using a reliable theory for gravitation, for
instance the general relativity of Einstein, to infer the properties
of the material content of the universe. This is naturally the approach of an observer who is unaware of internal dimensions. In fact, we may be living in a higher dimensional space which appears effectively three
dimensional since the internal dimensions are so small that
they evade direct and local detection. However, the internal
dimensions may still be controlling the dynamics of the external
dimensions that we observe. Hence, while we are interpreting the
cosmological data within the framework of four dimensional general
relativity, the components related to the internal dimensions and
the higher dimensional fluid we introduced could appear as an effective source in the 4-dimensional Einstein's field
equations. An observer who lives in four dimensions would use the 4-dimensional Einstein's field equations:
\begin{equation}
\label{eqn:EFEconv}
\tilde{R}_{ij}-\frac{1}{2}\tilde{R}\tilde{g}_{ij} =-{\tilde{k}}\tilde{T}_{ij},
\end{equation}
where $i$ and $j$ run through $0,1,2,3$ and ${\tilde{k}}=8\pi \tilde{G}$ with $\tilde{G}$ being the gravitational coupling constant. $\tilde{T}_{ij}$ is the energy-momentum tensor that will be inferred from the kinematics we obtain in our higher dimensional models. We know that the present universe is very well described by the spatially flat four dimensional Robertson-Walker metric. We do not expect the fluids to have bulk motion at cosmological scales, hence we assume that the fluid is at rest in the comoving coordinates. The Robertson-Walker metric admits only perfect fluid representation of the energy-momentum tensor as long as the fluid is at rest in the comoving coordinates, hence the energy-momentum tensor that we will infer can be represented by
\begin{equation}
\tilde{T}_{ij}=\text{diag}[-\tilde{\rho},\tilde{p},\tilde{p},\tilde{p}]
\end{equation}
where $\tilde{\rho}$ and $\tilde{p}$ are the induced energy density and pressure. Writing down the Einstein field equations in $(1+3)$-dimensions in the framework of conventional Robertson-Walker metric, we have
\begin{equation}
\label{eqn:EFE4Drho}
3\frac{\dot{a}^2}{a^2}+3\frac{\kappa_{a}}{a^2}={\tilde{k}}\tilde{\rho},
\end{equation}
\begin{eqnarray}
\label{eqn:EFE4Dp}
\frac{\dot{a}^2}{a^2}+2\frac{\ddot{a}}{a}+\frac{\kappa_{a}}{a^2}=-{\tilde{k}}\tilde{p}.
\end{eqnarray}
The observables in these equations are $\frac{\dot{a}}{a}$, $\frac{\ddot{a}}{a}$ and $\frac{\kappa_{a}}{a^2}$, which in return give us the opportunity to infer the effective energy density and pressure of the physical content of the universe. One may now observe how the components of the higher dimensional universe appear as an effective energy-momentum source in the four dimensional universe:
\begin{equation}
\label{eqn:EFE4Drhoint}
{\tilde{k}}\tilde{\rho}=k\rho_{0}-\Lambda-3n\frac{\dot{a}}{a}\frac{\dot{s}}{s}-\frac{1}{2}n(n-1)\frac{\kappa_{s}+\dot{s}^2}{s^2},
\end{equation}
\begin{equation}
\label{eqn:EFE4Dpint}
{\tilde{k}}\tilde{p}=kp+\Lambda + n\frac{\ddot{s}}{s}+2n\frac{\dot{a}}{a}\frac{\dot{s}}{s}+\frac{1}{2}n(n-1)\frac{\kappa_{s}+\dot{s}^2}{s^2},
\end{equation}
subject to the constraints (\ref{eqn:ansatz1}) and (\ref{eqn:ansatz2}).

We consider the solutions with the flat external space case such that $\kappa_{a}= 0$ and $\kappa_{s}\neq 0$. Using (\ref{eqn:afle}), (\ref{eqn:sfle}) and (\ref{eqn:pressurefle}) in (\ref{eqn:EFE4Drhoint}) and (\ref{eqn:EFE4Dpint}) the induced four dimensional effective energy density and pressure become
\begin{equation}
\tilde{\rho}=\frac{1}{{\tilde{k}}}\frac{2n}{3+n}(\Lambda-k\rho_{0})\left(\frac{s_{0}e^{-2\frac{\bar{n}}{n}\sqrt{\Lambda-k\rho_{0}} \; t}+\frac{n(n-1)}{8}\frac{\kappa_{s}}{s_{0}(\Lambda-k\rho_{0})}}{s_{0}e^{-2\frac{\bar{n}}{n}\sqrt{\Lambda-k\rho_{0}} \; t}-\frac{n(n-1)}{8}\frac{\kappa_{s}}{s_{0}(\Lambda-k\rho_{0})}}\right)^{2},
\end{equation}
\begin{equation}
\tilde{p}=-\frac{1}{{\tilde{k}}}\frac{2n}{3+n}(\Lambda-k\rho_{0})\left[1+\kappa_{s}\frac{(n+2)(n-1)}{2(\Lambda-k\rho_{0})}\frac{e^{-2\frac{\bar{n}}{n}\sqrt{\Lambda-k\rho_{0}} \; t}}{\left(s_{0}e^{-2\frac{\bar{n}}{n}\sqrt{\Lambda-k\rho_{0}} \; t}-\frac{n(n-1)}{8}\frac{\kappa_{s}}{s_{0}(\Lambda-k\rho_{0})}\right)^2}\right].
\end{equation}
Therefore, the induced EoS parameter for the fluid in the effective four dimensional universe would be
\begin{equation}
\tilde{w}\equiv\frac{\tilde{p}}{\tilde{\rho}}=-1-\kappa_{s}\frac{n-1}{\Lambda-k\rho_{0}}\frac{e^{-2\frac{\bar{n}}{n}\sqrt{\Lambda-k\rho_{0}} \; t}}{\left(s_{0}e^{-2\frac{\bar{n}}{n}\sqrt{\Lambda-k\rho_{0}} \; t}+\frac{n(n-1)}{8}\frac{\kappa_{s}}{s_{0}(\Lambda-k\rho_{0})}\right)^2}.
\end{equation}
The total mass within the 3-volume scale factor ($V_{3}=a^{3}$) of the effective universe reads
\begin{eqnarray}
\tilde{M}\equiv \tilde{\rho}V_{3}&=&\frac{2}{\tilde{k}}\frac{n}{3+n}(\Lambda-k\rho_{0}) \;\left( \frac{a_{0} e^{2\frac{\bar{n}}{3}\sqrt{\Lambda-k\rho_{0}} \; t}+\frac{3}{4}\frac{\kappa_{a}}{a_{0}(\Lambda-k\rho_{0})}}{a_{0}e^{2\frac{\bar{n}}{3}\sqrt{\Lambda-k\rho_{0}} \; t}-\frac{3}{4}\frac{\kappa_{a}}{a_{0}(\Lambda-k\rho_{0})}}\right)^2\\
\nonumber
&&\times\; V_{0}\left(s_{0}e^{-\frac{\bar{n}}{n}\sqrt{\Lambda-k\rho_{0}} \; t}-\frac{n(n-1)}{8}\frac{\kappa_{s}}{s_{0}(\Lambda-k\rho_{0})}e^{\frac{\bar{n}}{n}\sqrt{\Lambda-k\rho_{0}} \; t}\right)^{-n}.
\end{eqnarray}

On the other hand, for the special case $\Lambda-k\rho_{0}<0$ with $\kappa_{s}=1$, $n\geq 2$ and $s_{0}=\sqrt{\frac{n(n-1)}{8(k\rho_{0}-\Lambda)}}$ the induced energy density
\begin{equation}
\tilde{\rho}=\frac{2}{\tilde{k}}\frac{n}{3+n}(k\rho_{0}-\Lambda)\tan^{2}\left(\frac{\bar{n}}{n}\sqrt{k\rho_{0}-\Lambda}\;t \right),
\end{equation}
and the induced pressure
\begin{equation}
\tilde{p}=-\frac{2}{\tilde{k}}\frac{n}{3+n}(k\rho_{0}-\Lambda) \left[\tan^{2}\left(\frac{\bar{n}}{n}\sqrt{k\rho_{0}-\Lambda}\;t \right)-\frac{2}{n}\cos^{-2}\left(\frac{\bar{n}}{n}\sqrt{k\rho_{0}-\Lambda}\;t \right)\right].
\end{equation}
Therefore, the induced EoS parameter
\begin{equation}
\tilde{w}=-1+\frac{2}{n}\sin^{-2}\left(\frac{\bar{n}}{n}\sqrt{k\rho_{0}-\Lambda}\;t \right).
\end{equation}
The total mass within the 3-volume scale factor $V_{3}$
\begin{equation}
\tilde{M}=\frac{2}{\tilde{k}}\frac{n}{3+n}(k\rho_{0}-\Lambda)\left(\frac{1}{2}\frac{n(n-1)}{k\rho_{0}-\Lambda}\right)^{-\frac{n}{2}}V_{0}\left[\frac{\tan\left(\frac{\bar{n}}{n}\sqrt{k\rho_{0}-\Lambda}\;t \right)}{\cos^{\frac{n}{2}}\left(\frac{\bar{n}}{n}\sqrt{k\rho_{0}-\Lambda}\;t \right)}\right]^{2}.
\end{equation}

For consistency, we check the case where both the external and the internal spaces are flat, i.e., $\kappa_{a}=0=\kappa_{s}$. We obtain the induced energy density, pressure and the EoS parameter of the four dimensional effective fluid as follows:
\begin{equation}
\label{eqn:rrho}
\tilde{\rho}=\frac{2}{{\tilde{k}}}\frac{n}{3+n}(\Lambda-k\rho_{0}),
\end{equation}
\begin{equation}
\tilde{p}=-\frac{2}{{\tilde{k}}}\frac{n}{3+n}(\Lambda-k\rho_{0}),
\end{equation}
\begin{equation}
\tilde{w}=\frac{\tilde{p}}{\tilde{\rho}}=-1.
\end{equation}
The total mass within the 3-volume scale factor turns out to be
\begin{eqnarray}
\tilde{M}\equiv \tilde{\rho}a^3=\frac{2}{\tilde{k}}\frac{n}{3+n}(\Lambda-k\rho_{0}) {a_{0}}^{3} e^{\bar{n}\sqrt{\Lambda-k\rho_{0}} \; t}.
\end{eqnarray}
These expressions agree with the results given in Ref. \cite{DereliTucker83}.

\begin{figure}[ht]

\begin{minipage}[b]{0.49\linewidth}
\centering
\includegraphics[width=1\textwidth]{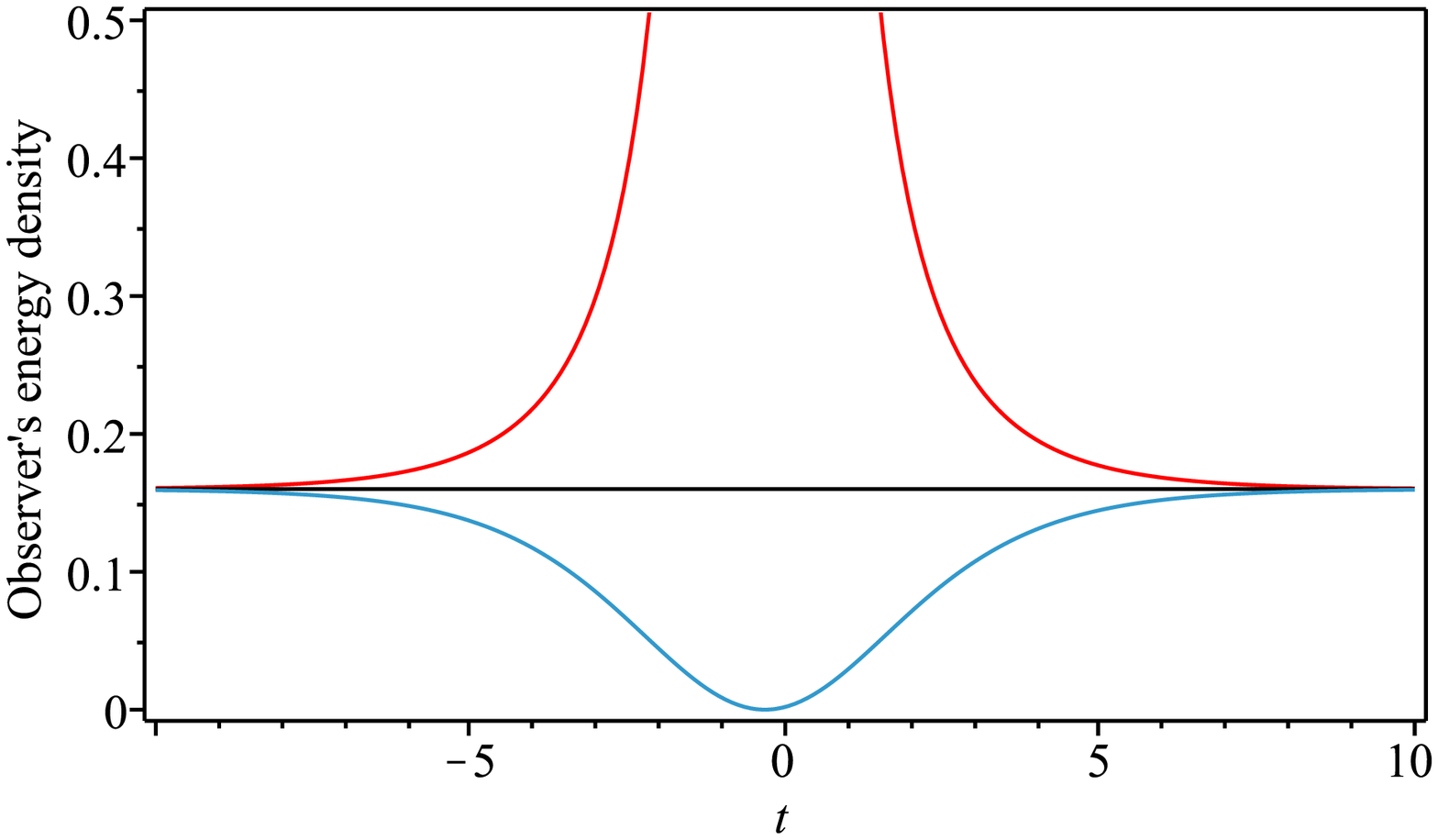}
\caption{The energy density would observer infer versus cosmic time $t$ for $n=2$. Blue, black and red curves represent spatially closed, flat and open internal spaces respectively.}
\label{fig:ha2}
\end{minipage}
\hspace{0.01\linewidth}
\begin{minipage}[b]{0.49\linewidth}
\centering
\includegraphics[width=1\textwidth]{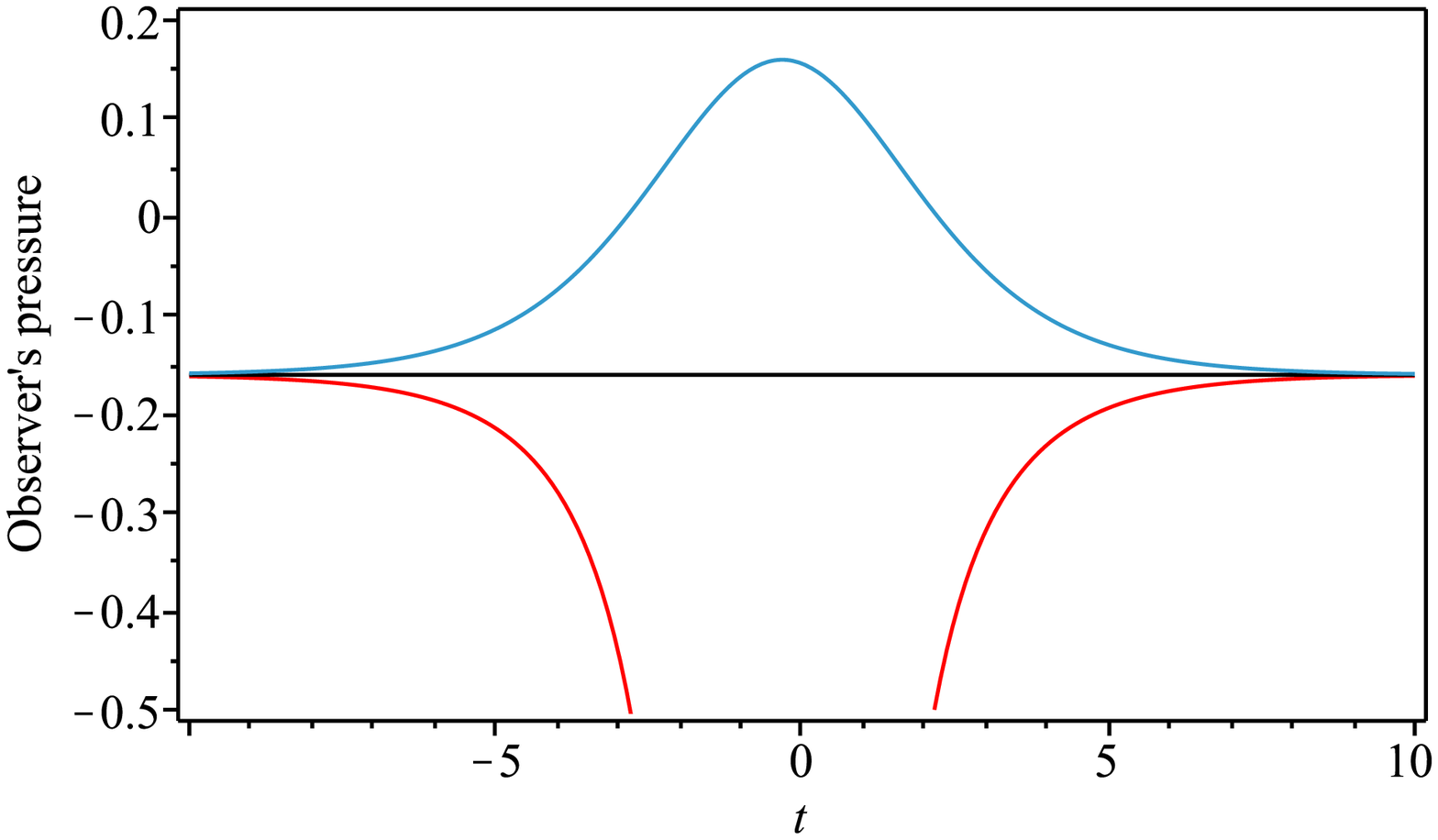}
\caption{The pressure of the fluid would observer infer versus cosmic time $t$ for $n=2$. Blue, black and red curves represent spatially closed, flat and open internal spaces respectively.}
\label{fig:pp2}
\end{minipage}

\begin{minipage}[b]{0.49\linewidth}
\centering
\includegraphics[width=1\textwidth]{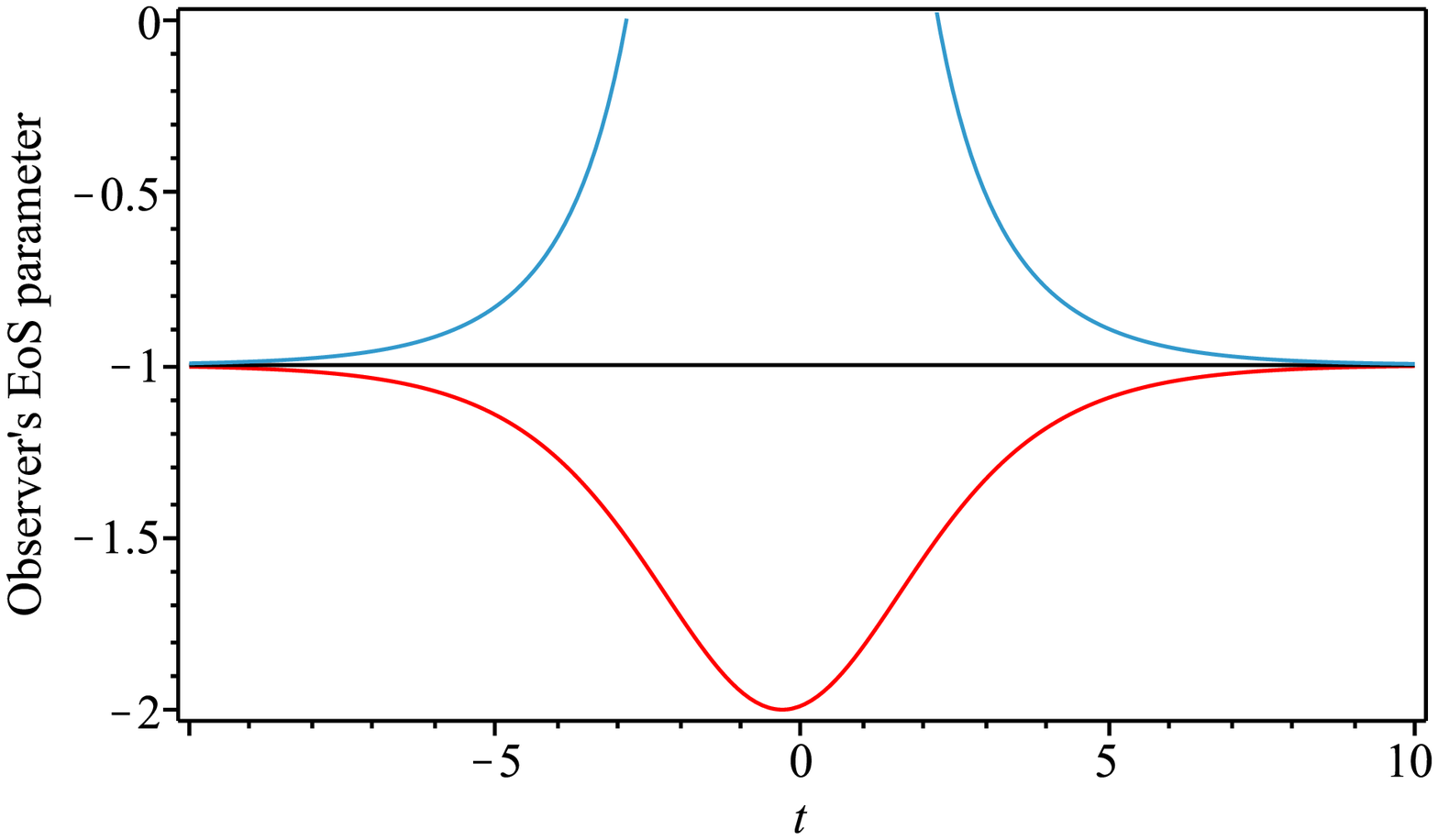}
\caption{The EoS parameter of the fluid would observer infer versus cosmic time $t$ for $n=2$. Blue, black and red curves represent spatially closed, flat and open internal spaces respectively.}
\label{fig:w22}
\end{minipage}
\hspace{0.01\linewidth}
\begin{minipage}[b]{0.49\linewidth}
\centering
\includegraphics[width=1\textwidth]{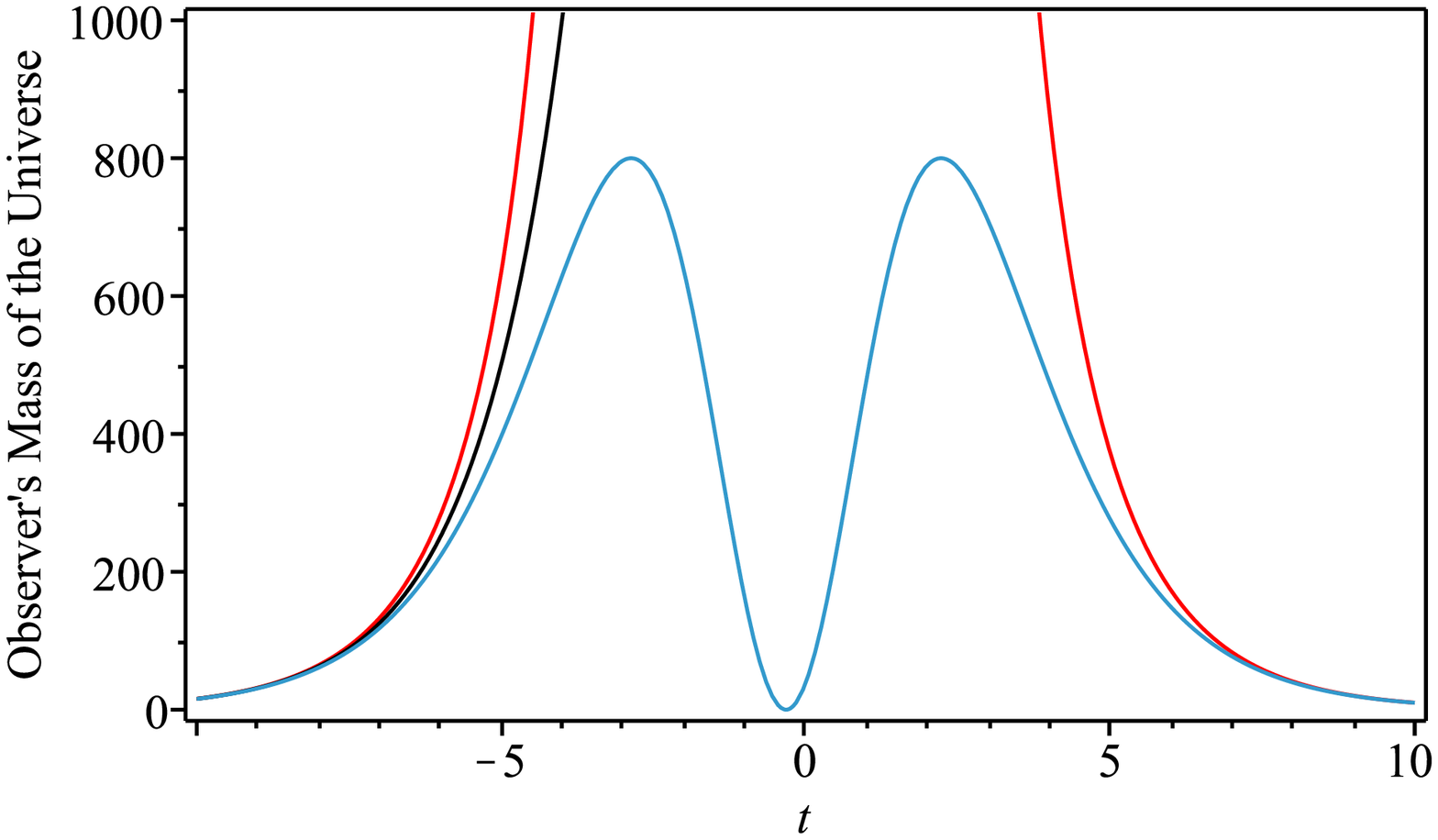}
\caption{Mass of the universe would observer infer versus cosmic time $t$ for $n=2$. Blue, black and red curves represent spatially closed, flat and open internal spaces respectively.}
\label{fig:M3}
\end{minipage}

\end{figure}

The effective energy density, pressure, EoS parameter and total mass within the 3-volume scale factor we obtained above are the ones that an observer would infer using the conventional four dimensional Einstein's field equations whenever the internal dimensions are so small that the observer is unaware of their presence. The energy density and pressure of the four dimensional effective fluid are dynamical for $\kappa_{s}\neq 0$ but both are constants for $\kappa_{s}=0$ and we give the generic behavior of the four dimensional effective fluid's energy density in Fig. \ref{fig:ha2}, pressure in Fig. \ref{fig:pp2} and EoS parameter in Fig. \ref{fig:w22} for the choice $n=2$. We would like to note that the behavior depicted in these figures corresponds to the behavior of the effective fluid that would generate the kinematics of the external space (solid curves) given in Fig \ref{fig:aet}. One may observe that the external space exhibits de Sitter expansion when the internal space is flat, and exhibits an almost de Sitter expansion for $t\ll t_{\rm c}$, but deviates from de Sitter expansion as the time approaches $t_{\rm c}$, in different ways for the cases with closed and open internal spaces.

In the case of closed internal space, the positive curvature speeds up the expansion rate of the external space to an increasingly super-exponential expansion rate as the internal space becomes smaller, and the size of the external space diverges at a finite time $t=t_{\rm c}$. This behavior is the one we know from the Big Rip cosmological models \cite{Caldwell03,Nesseris,Cai}, though in our model external space has an infinite past. Indeed, one may observe that the induced energy density also diverges at $t=t_{\rm c}$ in Fig. \ref{fig:ha2} as it would in the Big Rip models that are constructed in conventional general relativity. Hence, the observer would infer a source whose energy density increases as the universe expands and which is known as phantom field characterized by an EoS parameter less than $-1$ , is controlling the dynamics of the universe, as seen in Fig. \ref{fig:w22}. One may observe from Fig. \ref{fig:M3} that the mass within the volume scale factor also diverges at $t=t_{\rm c}$. In this case, the observer would think that matter is being created faster and faster as the universe expands, though in our model both the energy density and the volume in higher dimensions are constants.

In the case of open internal space, the negative curvature slows down the expansion rate of the external space more and more as the internal space becomes smaller and does not allow it to reach zero size but a non-zero minimum size is reached at $t=t_{\rm c}$. Hence the external space scale factor cannot evolve to infinitely large values but to a maximum value and then starts contracting. Meanwhile, the induced energy density remains almost constant for $t\ll t_{\rm c}$, starts to decrease considerably as $t$ approaches $t_{\rm c}$ and becomes zero at $t=t_{\rm c}$ (Fig. \ref{fig:ha2}). The induced mass of the physical universe increases almost exponentially up to a certain time, but this increase stops at some point and the induced mass starts to decrease and the universe evolves into an empty universe at $t=t_{\rm c}$ (Fig. \ref{fig:M3}). In this case, the observer would think that matter is being created slower and slower as the universe evolves and after some time the observer would think that matter has disappeared as the universe continues to evolve.

In the case of flat internal space, the observed universe expands exponentially and hence eternally. The induced energy density (Fig. \ref{fig:ha2}) is constant and the induced mass of the universe (Fig. \ref{fig:M3}) increases proportionally with the volume of the external space. This is a behavior of the well known four dimensional de Sitter solution of the Einstein's field equations with a positive cosmological constant. However, in our model the exponential expansion of the external space is not due to positive values of the Einstein's four dimensional cosmological constant but the positive values of the higher dimensional cosmological constant with anti-de Sitter sign.

The cosmological constant was first proposed by Einstein to obtain a static ($\dot{a}=0$) universe model. It was simply taken as the constant curvature of the empty spacetime and could take any real value. The positive cosmological constant is also mathematically equivalent to conventional vacuum energy $\tilde{\rho}_{\rm vac}=\tilde{k}\tilde{\Lambda}_{\rm E}$ with an EoS $\tilde{p}_{\rm vac}=-\tilde{\rho}_{\rm vac}$ in conventional four dimensional general relativity. This gave rise to one of the most significant controversies of modern physics that is known as the cosmological constant problem \cite{Zeldovich,Weinberg89,Sahni00}. Here we have a higher dimensional universe model and hence the higher dimensional vacuum energy can be thought as a constituent of our higher dimensional effective fluid $\rho_{0}$. Note that, assuming $k$ and $\tilde{k}$ of the same sign, any fluid with positive energy density will contribute negatively to the four dimensional effective fluid that is induced from the kinematics of the four dimensional external space. The higher dimensional negative cosmological constant, on the other hand, contributes positively. The higher dimensional $\Lambda$ we considered has anti-de Sitter sign and hence its positive values correspond to negative energy density with a positive pressure with an EoS parameter equal to $-1$. Accordingly, we can safely interpret it as the inherent curvature of the higher dimensional empty spacetime rather than vacuum energy density of the higher dimensional universe. Therefore, in our model, the energy of the vacuum is not the source that is responsible for the accelerating expansion of the external space but it is the higher dimensional negative cosmological constant. In fact, the higher dimensional vacuum energy would slow down in this case the expansion rate of the external space.

\section{Final remarks}
Cosmological models for which the higher dimensional volume is constant in time had already appeared in the literature in various contexts \cite{Freund82,DereliTucker83,BleyerZhuk96,RainerZhuk00,HoKephart10}. Here we made use of the constant higher dimensional volume assumption as a constraint that accompanies the constant higher dimensional energy density assumption. The constancy of the $(3+n)$-volume of the higher dimensional universe may be interpreted in two ways. Firstly it may be thought as the volume of the $(1+3+n)$-dimensional one and only universe. A constant and finite volume with constant energy (mass) density corresponds to a very familiar situation; namely an incompressible fluid. Hence, in the context of this interpretation, we may think that in our models the matter in $(3+n)$-dimensional space behaves like a $(3+n)$-dimensional incompressible fluid, which can be described by the energy-momentum tensor we obtained in the solutions, under the influence of $(1+3+n)$-dimensional Einstein's gravity with a cosmological constant. On the other hand, we may regard the higher dimensional constant volume as the volume of a $(3+n)$-dimensional region that is forced to stay constant inside a larger space. This is due to the fact $(1+3+n)$-dimensional effective fluid is under the influence of $(1+3+n)$-dimensional Einstein's gravity with a cosmological constant. This interpretation can be illuminated in analogy with gravitational fluctuations in the smooth background of a three dimensional space defined by the squashing of a spherical ball into an ellipsoid without changing its volume.

One may further ask whether it is possible to estimate the size of the $(1+3)$-dimensional space. The mean scale factor of the $(3+n)$-space can be defined by $l_{0}\equiv{V_{0}}^{\frac{1}{3+n}}=a^{\frac{3}{3+n}}s^{\frac{n}{3+n}}$, which depends on the number of internal dimensions as well as the length scales of the external and internal dimensions. Hence, in order to estimate $l_{0}$ we need to fix the length scales $a$, $s$ and the number of the internal dimensions $n$. We expect the length scale of $s$ to remain at unobservable sizes i.e. in order not to contradict observations we expect the size of the extra dimensions to be less than $\sim l_{\rm LHC}=10^{-20}$ m that corresponds to the TeV energy scale that is probed by the LHC (Large Hadron Collider). Hence we may carry out our analyses by assuming the size $s_{1}$ of the extra dimensions today is somewhere between the Planck length scale and LHC length scale, that is $10^{-35}{\rm{m}}\lesssim s_{1} \lesssim 10^{-20} {\rm{m}}$. On the other hand, the scale of the observed universe is $\sim 10^{27} {\rm{m}}$ (46 billion light years). Accordingly, the mean length scale of the $(3+n)$-dimensional space can be given in the range $10^{\frac{81-35n}{3+n}}{\rm{m}} \lesssim l_{0}\lesssim 10^{\frac{81-20n}{3+n}}{\rm{m}}$ for various numbers of internal dimensions. Considering this range, we give the mean length scale of the higher dimensional volume against the number of internal dimensions in Table \ref{tabular:lengthscales}. One may observe that, as the number of internal dimensions increases the mean length scale of the $(3+n)$-dimensional space approaches the length scale of the internal dimensions, i.e., $l_{0}\rightarrow s_{1}$ as $n\rightarrow\infty$. This behaviour leads to some interesting results. Because the mean length scale of the space is already constant, the internal dimensions are expected to remain almost constant with an arbitrarily large number of dimensions. Moreover, in the case of infinitely many internal dimensions, the size of the internal dimensions should be frozen (stabilize) at the magnitude $l_{0}$ as long as the size of the external space is not zero. Hence, for instance, if the internal dimensions today are at Planck scales, they should always remain at Planck scales.
\begin{table}[ht]\footnotesize
\begin{center}
    \begin{tabular}{ | c | c | c | c | }
    \hline
$\boldsymbol{n}$ & $\boldsymbol{3+n}$ & $\boldsymbol{l_{0}}$ for $s_{1}=l_{\textnormal{Planck}}$ and $a=10^{27}$ m & $\boldsymbol{l_{0}}$ for $s_{1}=l_{\textnormal{LHC}}$ and $a=10^{27}$ m \\ \hline
    1 & 4 & $10^{12} {\rm{m}}$ & $10^{15} {\rm{m}}$ \\ \hline
    2 & 5 & $10^{2} {\rm{m}}$ & $10^{8} {\rm{m}}$\\ \hline
    3 & 6 & $10^{-4} {\rm{m}}$ & $10^{4} {\rm{m}}$\\ \hline
    4 & 7 & $10^{-8} {\rm{m}}$ & $1 {\rm{m}}$\\ \hline
    5 & 8 & $10^{-12} {\rm{m}}$ & $10^{-2} {\rm{m}}$\\ \hline
    6 & 9 & $10^{-14} {\rm{m}}$ & $10^{-4} {\rm{m}}$ \\ \hline
    7 & 10 & $10^{-16} {\rm{m}}$ & $10^{-6} {\rm{m}}$ \\ \hline
    $\geq 8$ & $\geq 11$ & $ <10^{-16} {\rm{m}}$ & $<10^{-7} {\rm{m}}$\\ \hline
    $\infty$ & $\infty$ & $ l_{\textnormal{Planck}}=10^{-35}$ m & $l_{\textnormal{LHC}}=10^{-20} {\rm{m}}$ \\ \hline
    \end{tabular}
\end{center}
\caption{The length scale of the mean scale factor of the $3+n$ dimensional space $l_{0}$ according to the number of internal dimensions. We assume the length scale of the observed universe is $10^{27}$ m today and of the extra dimensions are somewhere between the Planck length scale $l_{\rm{Planck}}\sim 10^{-35}{\rm{m}}$ and LHC length scale, i.e., $l_{\rm{Planck}}\lesssim s_{1} \lesssim l_{\rm{LHC}}$. We calculate values of $l_{0}$ for $s_{1}=l_{\rm{Planck}}$ and $s_{1}=l_{\rm{LHC}}$.}
\label{tabular:lengthscales}
\end{table}

We have been arguing that an observer living in $(1+3)$-dimensions will be unaware of the presence of small internal dimensions. This is a valid assumption so long as the size of the internal dimensions remain within the range $l_{\rm{Planck}}\lesssim s \lesssim l_{\rm{LHC}}$. However,  even in this regime, the time variation in the size of the internal space would have important consequences for the fundamental constants of four-dimensional physics, for instance for the $(1+3)$-dimensional gravitational coupling $\tilde{k}=8\pi \tilde{G}$ in our study (See Ref. \cite{OverduinWesson97} and references therein for the possible effects of the extra dimensions on the four-dimensional physics). Such effects might be interpreted as an evidence for the presence of internal dimensions and used for providing information on the internal space by the observer. Therefore alternative to what we had done in the previous section, the 4-dimensional gravitational coupling $\tilde{k}=8\pi \tilde{G}$ could have been related to the higher dimensional gravitational coupling constant $k=8\pi G$ through the volume scale factor of the compact internal space as $\tilde{k}\propto k/{V_{3}}$. Then it follows that $\dot{\tilde{k}}/\tilde{k}=-nH_{s}$. However, this is not the approach we took in this paper. We included all the effects due to the internal space in the definition of the effective energy-momentum tensor $\tilde{T}_{ij}$ in four dimensions. A detailed discussion may be found in Ref. \cite{AkarsuDereli12}.

To summarize, we have discussed a higher dimensional cosmological model for which both the higher dimensional volume and energy density are constants. These assumptions lead to the interesting result that matter in our 3-space is neither created nor exhausted but redistributed between the internal and external spaces. Higher dimensional cosmological models with constant total volume that satisfy either power-law expansion \cite{Freund82,BleyerZhuk96,RainerZhuk00} or de Sitter expansion \cite{DereliTucker83} of the external space are already given in the literature. Here in this paper we have obtained various interesting new dynamics for the external space that yield a time varying deceleration parameter including oscillating cases when flat/curved external and curved/flat internal spaces are considered. Our model can be improved by considering modified gravity theories in higher dimensions rather than Einstein's general relativity as is done here.

\begin{center}
\textbf{Acknowledgments}
\end{center}
The research reported here is supported by a Post-Doc Research Grant by the Turkish Academy of Sciences (T{\"{U}}BA). \"{O}.A. also acknowledges the financial support from Ko\c{c} University and the support he received from the Abdus Salam International Center for Theoretical Physics (ICTP). \"{O}.A. is grateful for the hospitality of ICTP while the part of this research was being carried out. We also thank Anastasios Avgoustidis and Bar{\i}\c{s} \c{C}o\c{s}kun\"{u}zer for discussions.

\end{document}